\documentclass{article}

\usepackage{arxiv}

\usepackage[utf8]{inputenc} 
\usepackage[T1]{fontenc}    
\usepackage{hyperref}       
\usepackage{url}            
\usepackage{booktabs}       
\usepackage{amsfonts}       
\usepackage{nicefrac}       
\usepackage{microtype}      
\usepackage{lipsum}
\usepackage{bm}
\usepackage{xr}
\usepackage{graphicx}
\usepackage{float}
\usepackage{multirow}
\usepackage{amsmath} 
\usepackage{amssymb} 
\usepackage{amsfonts}
\usepackage{siunitx}
\usepackage{natbib} 

\usepackage{subfigure}

\title{Finding Continuity and Discontinuity in Fish Schools via Integrated Information Theory}

\author{
  Takayuki Niizato \\
  Faculty of Engineering, Information and Systems\\
  University of Tsukuba, Japan\\
  \texttt{t{\_}niizato@yahoo.co.jp} \\
   \And
 Kotaro Sakamoto\thanks{This author contributed equally to this work.} \\
  Human Biology Program \\
  University of Tsukuba, Japan \\
  \texttt{sakamoto@ccs.tsukuba.ac.jp} \\
  \AND
  Yoh-ichi Mototake \\
  Graduate School of Frontier Sciences \\
  University of Tokyo, Japan \\
  \And
  Hisashi Murakami \\
  Research Center for Advanced Science and Technology \\
  University of Tokyo, Japan \\
  \And
  Takenori Tomaru \\
  Department of Computer Science and Engineering \\ Toyohashi University of Technology, Japan \\
  \And
  Tomotaro Hoshika \\
  Faculty of Engineering, Information and Systems \\ University of Tsukuba, Japan \\
  \And
  Toshiki Fukushima \\
  Faculty of Engineering, Information and Systems \\ University of Tsukuba, Japan \\
}

\begin{document}
\maketitle

\begin{abstract}
Collective behaviour is known to be the result of diverse dynamics and is sometimes likened to a living system. Although many studies have revealed the dynamics of various collective behaviours, their main focus was on the information process inside the collective, not on the whole system itself. For example, the qualitative difference between two elements and three elements as a system has rarely been investigated. Tononi et al. have proposed Integrated Information Theory (IIT) to measure the degree of consciousness $\Phi$. IIT postulates that the amount of information loss caused by certain partitions is equivalent to the degree of information integration in the system. This measure is not only useful for estimating the degree of consciousness but can also be applied to more general network systems.
Here we applied IIT (in particular, IIT 3.0 using PyPhi) to analyse real fish schools ({\it Plecoglossus altivelis}). Our hypothesis in this study is a very simple one: a living system evolves to raise its $\Phi$ value. If we accept this hypothesis, IIT reveals the existence of continuous and discontinuous properties as group size varies. For example, leadership in the fish school emerged for a school size of four or above; but not below three. Furthermore, this transition was not observed by measuring mutual information or in a simple Boids model. This result suggests that integrated information $\Phi$ can reveal some inherent properties which cannot be observed using other measures. We also discuss how the fish recognition of the figure-ground relation, that is, what determines the relevant ON and OFF states, may reveal various optimal paths for obtaining the functional evolution of collective behaviour. 
\end{abstract}


\section{Introduction}
Collective behaviour, such as swarming \cite{Buhl_From_2006, Inherent_2009,Bazazi2012,  Attanasi_Finite_2014, Attanasi2014, Murakami2014}, fish schooling\cite{Ioannou2012, Strandburg-Peshkin2013, Berdahl2013, Murakami2015, Niizato2017} and bird flocking \cite{Ballerini2008, Cavagna2010, Cavagna2013, Bialek2014, Attanasi2015, Mora2016} has been widely observed in nature \cite{Couzin2007, Couzin2009, Sumpter_Collective_2010, Vicsek2012}. In some instances, individuals respond to the changing environment rapidly as one collective \cite{Cavagna2013, Bialek2014, Attanasi2015} and, in other cases, relatively good decision-making is achieved as a group \cite{Franks2003, Dyer2009, Bose2017}. Conflicts among individuals, as seen by an external observer, do not necessarily lead to group disruption; instead, they show the way to more an effective response as a group \cite{Couzin2011, Pinkoviezky2018}. The unity of this kind of animal behaviour remains one of the mysteries of nature \cite{Couzin2007}. 

Self-organised criticality (SOC) has been a good metaphor for interpreting these collective animal behaviours. If the group is in the intermediate state between order and disorder, it becomes possible to achieve enough flexibility and robustness as one system \cite{Bak1988,Tetzlaff2010, Niizato2012, Gunji2014, Gunji2014b, Niizato2018}. For example, the perturbations of flocks (or swarms) in SOC models optimise the effective correlation range of each bird and make it possible to accomplish fast information transfer \cite{Cavagna2010, Cavagna2013, Bialek2014, Attanasi2015}. However, when it comes to considering small groups, the same method cannot be applied, because, it is hard to assume that the interactions of individuals are homogeneous \cite{Herbert-Read2013, Jolles2017}. In particular, with regard to the subject of this study, it is conceivable that the interactions of two- and three-fish groups may be different \cite{Katz2011, Gautrais2012}. Many researchers, therefore, have considered information transfer (or causal relationships) among individuals in small groups\cite{Staniek2008, Butail2016,  Crosato2018}. The (local) transfer entropy is the preferred measure to use in this case\cite{Lizier2008, Lizier2012,Sun2014,James2016, Tomaru2016, YagmurErten2017}. For example, Crosato et al. \cite{Crosato2018} showed that the transfer of misinformation happens in five-fish school when the whole school changes direction. Other studies suggest that active information storage can predict the timing when nontrivial information transfer happens \cite{Wang2012, Wang2011}. Although the latter approaches also promise to give us a tremendous amount of information about what is happening in the group, they will not give us the information about what the system of collective behaviour is \cite{Albantakis2015}. The SOC approach certainly captures some aspects of what the system of collective behaviour is, but it gives little information about the causal structures inside the groups.

Before we go into detail about the difference between what is happening and what the system is, we need to introduce the concept of integrated information theory (IIT). IIT, which Tononi and other researchers have proposed, has been a rapidly developing area over the last two decades \cite{Balduzzi2009, Tononi2010, Barrett2011, Oizumi2014, Oizumi2015, Oizumi2016, Mayner2018}. The original aim was to estimate the degree of consciousness from brain activity \cite{Balduzzi2009, Tononi2010}. Recent studies suggest that IIT can capture and discriminate between various states of lost consciousness, such as dreamless sleep \cite{Massimini2005}, general anaesthesia \cite{Alkire2008} or vegetative states \cite{Gosseries2014}. Although IIT has several versions, its core concept is the same in principle, that is, the integrated information ($\Phi$) is defined as the degree of information loss caused by a certain partition of the system \cite{Tononi2010, Oizumi2016}(in the case of Barrett and Seth's version of IIT, $\Phi$ is the degree of the increase of uncertainty caused by a certain partition\cite{Barrett2011}. A computational comparison of many versions of IIT has been made by Mediano \cite{Mediano2018}). It is worth noting that Ito \cite{Ito2018, Ito2018b} has pointed out the fact that there are some intimate relations between the second law of information thermodynamics and IIT in terms of a projection onto a local reversible manifold. These structural resemblances suggest the possibility of unifying the concept of non-equilibrium thermodynamics and IIT. 

The key concept of IIT is that the whole cannot be reduced into its separated parts because the lost information would contain synergetic information produced by those parts. In this respect, the concept of IIT resonates with that of complex systems \cite{Bertalanffy1969}, for which the statement ''the whole is more than the sum of its parts'' has long been a slogan \cite{Hooker2011}. Since the intrinsic causal structures make the system irreducible into its parts, the integrated information (or $\Phi$) also can be a measure of the degree of wholeness as a single autonomous system\cite{Farnsworth2018}. 

There have been some applications of IIT to cellular automaton\cite{Albantakis2015}, animat \cite{Edlund2011} and Boolean networks \cite{Marshall2017}. For example, Albantakis et al. \cite{Albantakis2015} showed that average $\Phi$ values for 5 to 6 cells correlated well with their complexity, such as class III and IV, despite the very small number of cell sets. (The behaviours of 5 and 6 cells can be hardly discriminated on the basis of the behaviours of their constituent cells and, in general, the behaviours of small numbers of cellular automata are very similar to an external observer.) They also showed that all rules of class IV have all orders of concepts (i.e. irreducible subsets in the system) unlike other classes. 

The example of cellular automata illuminates the meaning of intrinsic properties for IIT. IIT reveals the differences among systems arising from different intrinsic causal structures (rules), rather than considering differences based on external behaviour. That is why we said previous approaches (especially, transfer entropy) captured not what the system is but what is happening. Now we can ask the following question: What is the difference of collective behaviour in terms of the intrinsic causal structure perspective? In this paper, we ask the following: Does the number of agents in a system make its intrinsic properties different? In other words, if the group size changes, what remains the same (continuous) and what changes (discontinuous) in the group? Also, are any new factors introduced, which were not present before? This kind of question is rarely asked in animal collective behaviour, but one study suggests that schools of three fish and schools of two fish have different kinds of interactions \cite{Katz2011, Gautrais2012}. Another suggests that the search strategies of fish in groups of different sizes are essentially different when they are in an unfamiliar environment \cite{Niizato2017}. However, all these studies constrain the number of individuals in the group to three or less and their methods are difficult to generalise to larger groups. Furthermore, these methods never indicate any differences in terms of the group’s intrinsic causal structure.

In this paper, we apply IIT (in particular, IIT 3.0 using PyPhi \cite{Oizumi2014, Mayner2018}) to schools of two to five fish ({\it Plecoglossus altivelis}) and show the intrinsic differences between these groups. To apply IIT to the collective behaviour of animals, we propose a simple hypothesis, namely, that a living system evolves to raise its integrated information. This hypothesis is not a peculiar one because some studies have suggested that, for some artificial systems selected by their fitness, $\Phi$ values were correlated with fitness \cite{Albantakis2015}. Thus, to raise $\Phi$ means to raise fitness in a given environment. Adopting this hypothesis, we found that there is a kind of continuity and discontinuity with respect to school size. The main finding is that there is a discontinuity between three- and four-fish schools, which is a difference that hasn't received a lot of attention previously. Interestingly, the difference between these two systems corresponds to the existence of leadership (more precisely, reducing the field of view for fish’s recognition introduces the existence of the leadership). Furthermore, our results are never replicated by a Boids-type model for the same conditions.

\section{Results}
\label{sec:Results}

\subsection{Definition to apply IIT to fish schools}

\begin{figure}[ht]
\centering
\includegraphics[width=\linewidth]{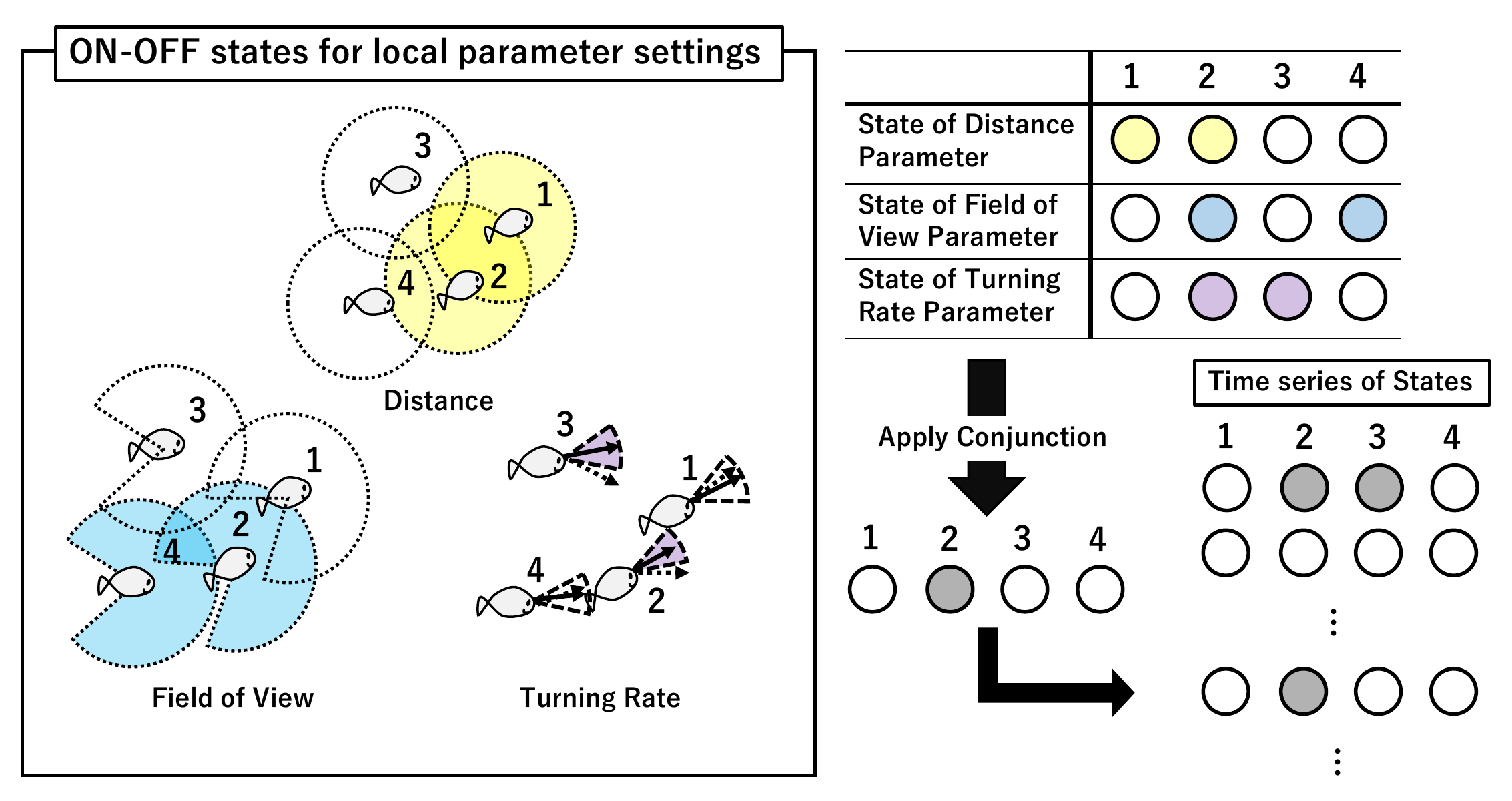}
\caption{The definition of ON and OFF states for local parameter settings. Three parameters determine a school's state (Yellow: distance, Blue: Field of View, Purple: Turning rate). Coloured individuals are in the ON state. We take a conjunction of the three school states to obtain the final school state at time $t$. Then we compute $\Phi$ from a time series of these states by using PyPhi.}
\label{Figure_1}
\end{figure}

To apply IIT 3.0, we define ON and OFF states of an individual in a fish school. In this paper, the ON state means some interaction would occur in a given context. For example, if two individuals are within a certain radius, the state of both individuals are ON (some information transfer would occur between them). This is a symmetric interaction. In the same way, we consider two other interactions to define ON and OFF states for fish in the school: visual field and turning rate interactions (see Fig. \ref{Figure_1}). A visual field interaction means the individual is in the ON state when some other agents are within its visual field. This allows us to consider asymmetrical relations in contrast to the symmetric distance condition. The turning rate interaction is one in which a direction change above a certain value puts the individual in the ON state. This ON state transfers information to other agents in the next time step, so the interaction between individuals is a delayed one. The direction changing rate is a very important measure for collective behaviour, empirically and theoretically \cite{Couzin2002, Vicsek2012, Strandburg-Peshkin2013}. 

In this paper, we assume a fish always evaluates these three kinds of information simultaneously. So, we take conjugation (i.e. AND) of the obtained 3 bits of information (for instance, IF Distance:ON, Visual field:ON, Turning rate:OFF, THEN state OFF) to produce an overall state for a fish. Applying the same process to each fish at a time $t$, we obtain the time series of the states of the $n$-fish. Then we can compute $\Phi$ and other values (the number of concept) from the obtained time series. One time step, in this paper, is defined as 0.05 (0.10, 0.20) s. This value roughly corresponds to the fish's reaction timescale \cite{Crosato2018}.

To compute $\Phi$, we also define the network structure in the school. In this paper, we postulate the completely connected network not including self-loops. This assumption comes from the experimental fact that each fish has some contact with (or falls within the visual field of) all individuals in the group during the long series of recorded events (10-15 min). Therefore, it is natural to assume that some interactions happened among all members (In Table S1, we give the minimal distance throughout the events. The data shows all fish have a contact within 5mm). 

Before we go into detail about our analysis, it is necessary to understand what the states ON and OFF mean for the fish. Biological information systems, such as the brain have an explicit ON state, that is, firing neurons. In contrast, the ON state for each fish is its recognition of a certain environment, that is, it is the state of a characteristic factor to which each fish pays attention. Since there are various kinds of information to take into account, there is no explicit ON state in fish school.  (This kind of ambiguity is not a demerit of our analysis. We will come back this issue in the Discussion.)

\subsection{$\Phi$ values for local parameter settings}
First, we confirmed the fact that $\Phi$ increases with group size (from two to five fish) on average. This trend is also observed in the Boids model (with the same parameter setting, see Table S2) but values are higher than those for real fish schools. This result is a very natural one because the degree of integrity becomes high when each agent keeps their distance almost constant and moves as one collective throughout the series of events. Compared with the Boids model, fish in the real fish school connect more loosely with each other. As a result, $\Phi$ for real fish schools is smaller than in models.

\begin{figure}[H]
\centering
\includegraphics[width=\linewidth]{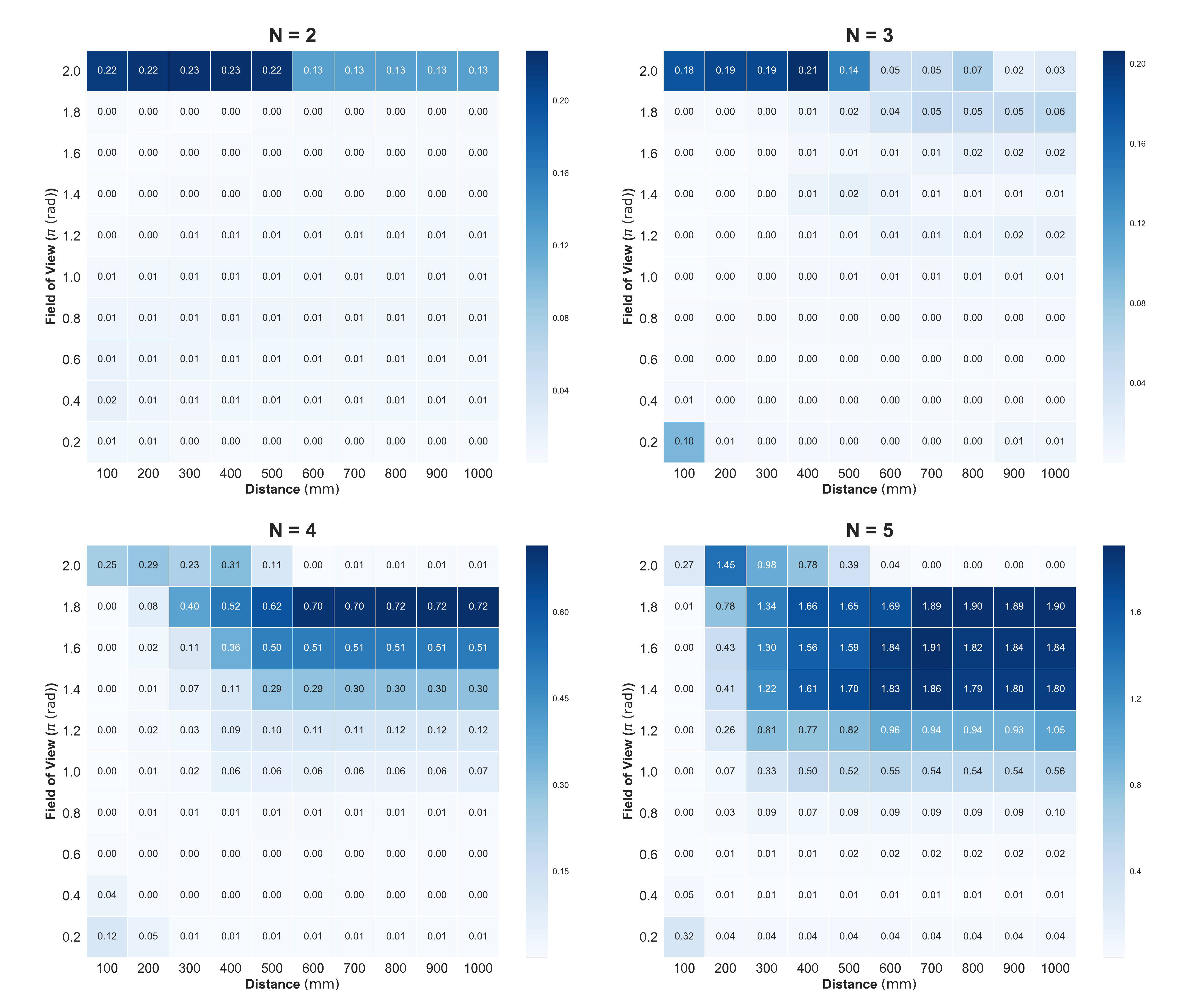}
\caption{The heat map shows $\Phi$ values for distance (horizontal axis) and field of view (vertical axis) parameter values. The value of $\Phi$ for two- and three-fish schools depends only on distance; whereas it depends on both parameters in four- and five-fish schools. The time step is $\Delta t = 0.05$ s. All other cases are described in the Supporting Information.}
\label{Figure_2}
\end{figure}

Fig. \ref{Figure_2}  shows that a qualitative change occurs when group size increases from three to four. Apparently, $\Phi$ values in two- and three-fish groups depend only on the distance threshold and not on the visual field. It appears that the leadership relation is not so important for fish groups smaller than 4 (an enlarged version of Fig. \ref{Figure_2} is given as \ref{fig:sfig31}). Leadership emerges when the group size is four or more. Interestingly, this trend is not observed in the Boids model and in the mutual information model. (See Figs. \ref{sfig1} and \ref{fig:sfig22a}. Fig. \ref{fig:sfig32} shows other parameter settings.) Furthermore, if we take a time step of 0.1 and 0.2 s instead of 0.05 s, the same tendency observed in almost cases (from Figs.\ref{fig:sfig33} to \ref{fig:sfig33-2}). Two-fish groups show high values around Field of View = $(1/5) {\pi}$ ({\it rad}) and Distance=200({\it mm}) (Fig. \ref{fig:sfig33} and \ref{fig:sfig33-2}). The leadership relation that emerges is, however, essentially different from that in large groups. This kind of behaviour may be called ``followership'' because the very narrow visual fields lead individuals to target the fish swimming ahead of him. 

\begin{figure}[H]
\centering
\includegraphics[width=\linewidth]{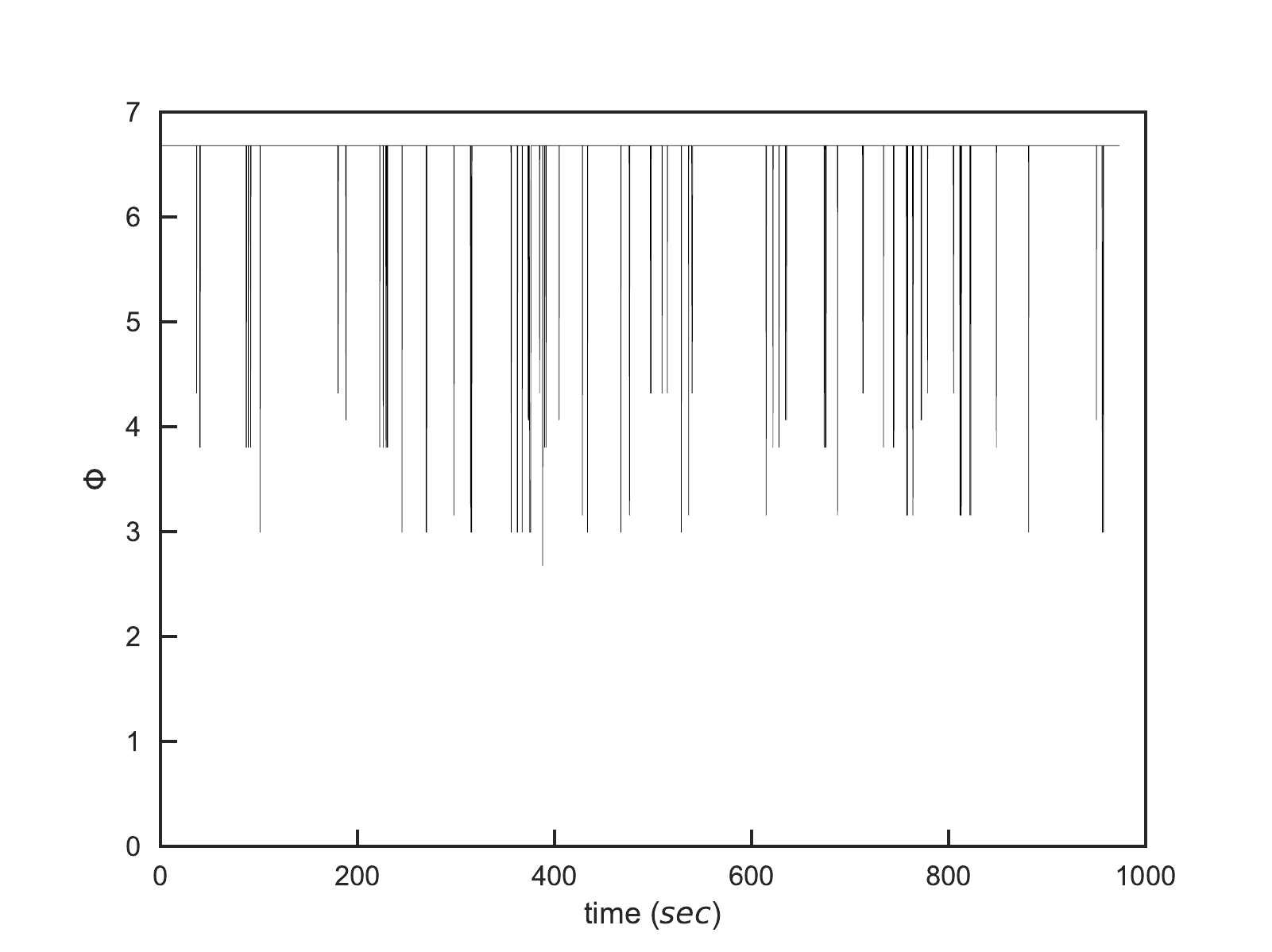}
\caption{An example of a time series of $\Phi$ obtained from real fish data ($N=5$). The reduction of $\Phi$ at various points means that leadership emerges in the group. Average $\Phi$ becomes higher when the group has a leader. Note that the average $\Phi$ in this figure is averaged from $\Phi$ values of 32 states in this case. The local parameter setting is Distance= 1000 ({\it mm}), Field of View = 6.02 ({\it rad}) Turning rate = 0 ({\it rad}), respectively.}
\label{Figure_3}
\end{figure}

Fig. \ref{Figure_3} is an example of a time series of $\Phi$. The abrupt reductions of $\Phi$ values correspond to the emergence of leadership. In IIT 3.0, the emergence of leadership never raises the $\Phi$ value; it always decreases it. Leadership decreases the integrity of the school because, if we cut between a leader and its followers, the integrity of the whole will be disrupted. So, the emergence of leadership itself raises $\Phi$ values on average (the highest $\Phi$ value corresponds when all fish are in the ON state); however, they decrease $\Phi$ values as a single state.

We also find that the turning rate is not so important for determining $\Phi$ values for cases with a short timescale ($\Delta t = 0.05$ s). However, turning rate becomes important information for long timescale events (see Fig. \ref{fig:sfig33}  and \ref{fig:sfig33-2}). Over a short timescale, relative positional information seems the most important for raising $\Phi$ values. 

The other intriguing measure is the number of concepts. Concepts are one of the critical notions in IIT 3.0 because $\Phi$ values are determined by their distribution in a conceptual space. A concept is, in short, the ability of ``difference makes difference'' as a subsystem. (Further explanation is given in the''Integrated information $\Phi$'' and ''Concept'' sections in the Supporting Information). If a system contains many concepts (up to $2^n - 1$ concepts exist for $n$ elements), that system has many irreducible components (i.e. it cannot be decomposed into its parts) as subsystems. The importance of the number of concepts can be observed in elementary cellular automata. The rule which shows class IV behaviour has all orders of concept unlike other classes \cite{Albantakis2015}. 

We found that there are areas that are rich in concepts despite low $\Phi$ values (Figs.S10 and S11). Combined with the results shown Fig. \ref{Figure_2}, we can distinguish three types of combination, that is, low $\Phi$ and few concepts, high $\Phi$ and many concepts, and low $\Phi$ and many concepts (there are no examples of the combination of high $\Phi$ and few concepts in our study). The most interesting case is the combination of low $\Phi$ and many concepts. These areas tend to get high $\Phi$ values if the number of fish increases. This observation suggests that low $\Phi$ values and many concepts provides the possibility of evolution if the condition (or environment) changes. (We also examined other measures. See \ref{fig:sfig35} to Fig. \ref{fig:sfig36}.)

\subsection{$\Phi$  values for global parameter settings}
\begin{figure}[H]
\centering
\includegraphics[width=\linewidth]{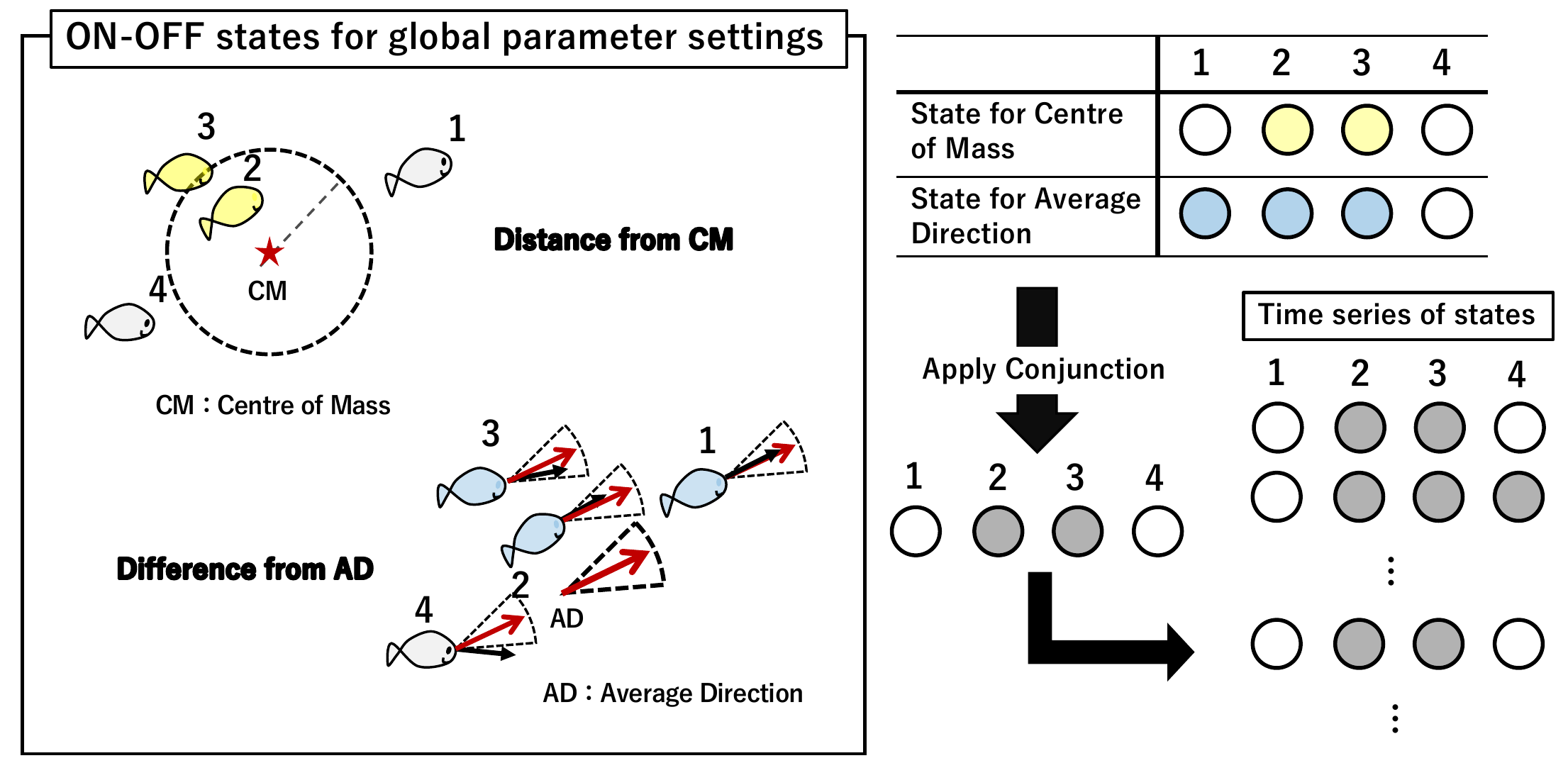}
\caption{The definition of ON and OFF states for global parameter settings. Two parameters determine a school's state (Yellow: Centre of Mass, Blue: Average Direction). Coloured individuals are in the ON state. We take a conjunction of the two school states and obtain the final school state at time $t$. Then we compute $\Phi$ from a time series of these states by using PyPhi.}
\label{Figure_4}
\end{figure}

Next, we defined the ON and OFF states globally rather locally. That is, the states are determined by global measures of interaction rather than local ones, as previously. For this, we considered the average direction and the centre of mass. When the difference between a fish's direction and the average direction of the fish school is within a certain specified value, its state is ON (see Fig. \ref{Figure_4}). Similarly, when each fish's distance from the centre of mass of the school is smaller than a certain specified value, its state is ON. The main difference from the previous state definition is that these parameters require the existence of a single group to be postulated {\it a priori}. These values will make no sense if the group is divided into two groups. (It is possible that two independent coherent groups will be incoherent when considered as a whole.)

\begin{figure}[H]
\centering
\includegraphics[width=.6\linewidth]{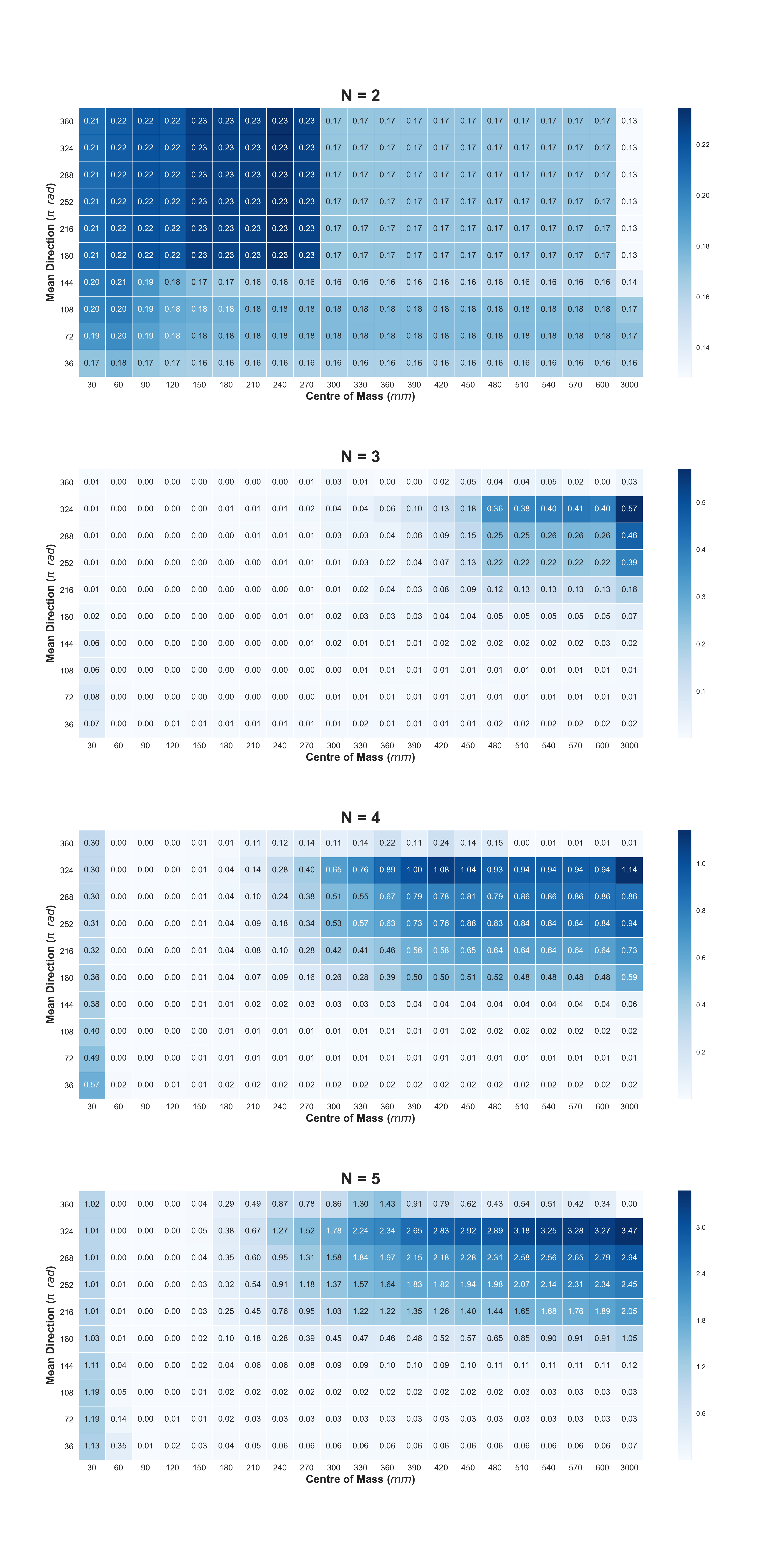}
\caption{The heat map shows $\Phi$ values with distance from the centre of mass (horizontal axis) and the difference from the average direction (vertical axis). Only two-fish schools show a different distribution. When the group size is above three, the distribution of $\Phi$ values becomes wider. The time step is $\Delta t = 0.05$ s.}
\label{Figure_5}
\end{figure}

As in the local case, $\Phi$ values also rise with group size (Fig. \ref{Figure_5}). This tendency is also observed in the Boids model. (Note, in particular, that the distribution of two-fish groups in the Boids model is very different from that of real two-fish schooling. See Fig. \ref{fig:sfig22b}). The main difference between local and global measures is the discontinuity occurs at a different point, that is, between two fish and three fish. The discontinuity between three- and four-fish schools is never observed for the global parameters. In this sense, three- and four-fish schools are continuous with respect to the global parameters.

\section{Discussion}

In this study, we applied IIT 3.0 to real fish schools and compared the results with those for another measure (mutual information) and another model (Boids) under the same conditions. Our results suggest the degree of integration $\Phi$ might pick up some unique information about real fish schools. From the $\Phi$ distributions derived with a certain set of parameters, we found a discontinuity between three- and four-fish school with local parameter setting but continuity with global setting: the recognition of leadership raises the degree of integrity above four but not below three. Changing the timescale from 0.05s to 0.2s, we confirmed the emergence of ``followership'' rather than leadership in two-fish groups (Fig. \ref{fig:sfig33} and \ref{fig:sfig33-2}). 
Therefore, their intrinsic causal structures are clearly distinct in terms of IIT, although two- and four-fish schools may exhibit leadership as a group.

This result is consistent with Albantakis's argument that IIT captures ``what a dynamical system is from its own intrinsic perspective'' (or ``how much and in which way it exists for itself, independent of an external observer'') rather than ``what is happening in a system from extrinsic perspective of an observer''\cite{Albantakis2015}. Along the lines of this statement, we can say the emergence of leadership represents what the system of a fish school is with respect to its group size. It is worth noting that IIT discriminates between three- and four-fish groups, which is a comparison that is rarely considered in the context of collective animal behaviour, although there are some studies that suggest a difference between two- and three-fish groups in terms of each fish's interactions with others (i.e. a difference in what is happening in the system from an extrinsic perspective') \cite{Katz2011, Gautrais2012, Niizato2017}.

Finally, we comment on the relation between animal recognition systems and the evolution of collective animal behaviour. In this paper, we have hypothesised that living systems evolve to raise their $\Phi$ value. This hypothesis itself is not a peculiar one because some studies have shown that the fitness of artificial systems, such as Animats and genetic Boolean networks, is correlated with $\Phi$ \cite{Edlund2011,Albantakis2015, Marshall2017}. Simple biological systems also show some connections between their functional units and $\Phi$ values (or their concepts) \cite{Marshall2017}. For example, in our study, the emergence of leadership in groups of four fish or more means each individual chooses to reduce its field of view in the group to raise $\Phi$ values (Fig.~\ref{Figure_2} shows the peak of $\Phi$ values of a five-fish group, indeed, shifting to make the field of view smaller than that of a four-fish group).

In our analysis, the factor which determines what is ON and OFF is a fish's recognition of its environment. In contrast with brain systems, ON states are dominant in a fish school. This fact means the OFF states are more informative than the ON states. The ON states, especially for local parameter settings, are important because the all-ON state for a school means all fish recognise that they are part of the same group. That is why the state of leadership (one fish in the group is the OFF state) reduces $\Phi$. 

We have confirmed that leadership never raises $\Phi$ when the group size is three or less. This fact indicates the two- and three-fish groups tend to show fission-fusion behaviour rather than leadership. In addition to this, three-fish schools can be said to be a kind of tipping point from a local to a global collective. From the view of the local perspective (local parameter settings), there seems to be no advantage for the group when a two-fish school becomes a three-fish school because $\Phi$ values never rise in this condition. On the other hand, from the global perspective (global parameter settings), increasing the group size from two to three means increasing $\Phi$ values. Therefore, the recognition of what is ON or OFF in those systems would change the $\Phi$ values radically and help the group to find its way to other optimal states of $\Phi$ values for other recognition. Our results suggest the evolution of real autonomic systems would become possible through IIT.

In this study, we avoided going deeply into the problem of timescale (we only used a relatively small timescale, which is roughly equal to a general fish's reaction time). Over longer timescales, other patterns of continuity and discontinuity may be found. Increasing the number of individuals may also give other results. However, the present practical computational limit of IIT 3.0 is around 7 or 8 individuals/neurons \cite{Mayner2018}, so some approximations will be needed to implement further analysis. Another area we didn't address is network structure. We supposed an all-connected network without self-loops in this paper because all fish came into contact with each other throughout the event. This will not always be true for large groups. Furthermore, some studies suggests that the network structure of real schools of fish is radically different from the Boids model one, and that they make a stable network called the $\alpha$-lattice \cite{Olfati-Saber2006,Olfati-Saber2007}. This type of network may prevent $\Phi$-raising trends observed in the Boids model.

\section{Methods}
\subsection{Ethics statement}

This study was carried out in strict accordance with the recommendations in the Guide for the Care and Use of Laboratory Animals of the National Institutes of Health. The protocol was approved by the Committee on the Ethics of Animal Experiments of the University of Tsukuba (Permit Number: 14-386). All efforts were made to minimize suffering.

\subsection{$\Phi$ computation}
All computations, in this paper, were performed using the PyPhi software package with the CUT{\_}ONE{\_}APPROXIMATION to $\Phi$. 

\subsection{Experimental Settings}
We studied {\it ayus} ({\it {\it Plecoglossus altivelis}}), also known as sweetfish, which live throughout Japan and are widely farmed in Japan. Juvenile {\it ayus} (approximately 7-14 {\it cm} in body length) display typical schooling behaviour, though adult {\it ayus} tend to show territorial behaviour in environments where fish density is low. We purchased juveniles from Tarumiyoushoku (Kasumigaura, Ibaraki, Japan) and housed them in a controlled laboratory. Approximately 150 fish lived in a 0.8 $m^3$ tank of continuously filtered and recycled fresh water with a temperature maintained at 16.4${}^\circ$C, and were fed commercial food pellets. Immediately before each experiment was conducted, randomly chosen fish were separated to form a school of each size and were moved to an experimental arena without pre-training. The experimental arena consisted of a $3{\times}3 m^2$ shallow white tank. The water depth was approximately 15 $cm$ so that schools would be approximately 2D. The fish were recorded with an overhead grey-scale video camera (Library GE 60; Library Co. Ltd., Tokyo, Japan) at a spatial resolution of 640 $\times$480 pixels and a temporal resolution of 120 frames per second.

\subsection{The definition of ON and OFF state for each parameter}
We define a function for each parameter that returns either 0 (OFF) or 1 (ON) for given input values. Generally, we denote a function as $F_{i}^{t}(\cdot)$, where $F$ is the name of the function, $i$ is the index of the individual and $t$ is the time. The arguments of the function can be either in the position vectors $\bm{x}_i(t)$ or the velocity vectors $\bm{v}_i(t)$ of each individual at time $t$. In general, the dimensions of these vectors are $d\leq 3$; the experimental setup used here gives $d=2$. The number of individuals is $n$.

\subsubsection{Local parameters}
\begin{itemize}
\item Distance function $D_{i}^{t}(\bm{x}_{1}(t), \bm{x}_{2}(t), \cdots, \bm{x}_{n}(t))$: $\mathbb{R}^{d} \times \mathbb{R}^{d} \times \cdots \times \mathbb{R}^{d} \xrightarrow{} \{ 0, 1 \}$

For each individual $i$ we obtain a set $S_{i}^{t}= \{j | d(\bm{x}_{i}(t), \bm{x}_{j}(t)) < \zeta, j \neq i  \}$  of all other individuals within a specified distance $\zeta$. Here $d(\bm{x}, \bm{y})$ gives the Euclidean distance between $\bm{x}$ and $\bm{y}$. Then, $D_{i}^{t}(\bm{x}_{1}(t), \bm{x}_{2}(t), \ldots, \bm{x}_{n}(t)) = 1$ when $|S_{i}^{t}|>0$ and is 0 otherwise, where $|S|$ denotes the number of elements of a set $S$. 
 
 \item Blind sight function $B_{i}^{t}(\bm{v}_{1}(t), \bm{v}_{2}(t), \cdots, \bm{v}_{n}(t)):\mathbb{R}^{d} \times \mathbb{R}^{d} \times \cdots \times \mathbb{R}^{d} \xrightarrow{} \{ 0, 1 \}$
 
 For each individual we form the set $O_{i}^{t} = \{j|$ arg($\bm{v}_{i}(t)$, $\bm{v}_{j}(t)$) $< \eta$, $j \neq i \}$ of all other individuals whose velocity vectors point in a direction within an angle $\eta$ of that of the focal individual. The function arg($\bm{v}_{1}(t)$, $\bm{v}_{2}(t)$) gives the angle between two vectors. Then, $B_{i}^{t}(\bm{v}_{1}(t), \bm{v}_{2}(t), \cdots, \bm{v}_{n}(t)) = 1$ when $|O_{i}^{t}| > 0$ and is 0 otherwise. 

 \item Turning rate function $T_{i}^{t}(\bm{v}_{i}(t), \bm{v}_{i}(t-\Delta t)):\mathbb{R}^{d} \times \mathbb{R}^{d} \xrightarrow{} \{ 0, 1 \}$

The turning rate function returns 1 when an individual's turning rate exceeds a specified threshold$ \delta$. That is, $T_{i}^{t}(\bm{v}_{i}(t), \bm{v}_{i}(t-\Delta t))= 1$ when arg($\bm{v}_{i}(t)$, $\bm{v}_{i}(t-\Delta t)) \geq \delta$ and is 0 otherwise. The time step used in this paper is $\Delta t = 0.05$, $\Delta t = 0.1$ or $\Delta t = 0.2$ s.

  To obtain the states of the fish school, we take a conjunction of these result, that is, $D_{i}^{t}(\bm{x}_{1}(t), \bm{x}_{2}(t), \cdots, \bm{x}_{n}(t)) \wedge  B_{i}^{t}(\bm{v}_{1}(t), \bm{v}_{2}(t), \cdots, \bm{v}_{n}(t)) \wedge T_{i}^{t}(\bm{v}_{i}(t), \bm{v}_{i}(t-\Delta t))$ for each individual $i$. The conjunction is given as $\wedge : \{ 0, 1 \}^2 \xrightarrow{} \{ 0, 1 \}$ where $1 \wedge 1 = 1$ and is  0 otherwise. Thus the state of  each individual $i$ at time $t$ is $s_{i}(t; \zeta,\eta, \delta) \in \{0,1\}$ which depends on the triplet of parameter values $(\zeta,\eta, \delta)$. The state of the school at time $t$ is then a vector $\bm{s}(t) = (s_1(t), s_2(t), \ldots, s_n(t)) \in \{0,1\}^n$, where the parameter dependence has been omitted for simplicity.

\end{itemize}

\subsubsection{Global parameters}
\begin{itemize}

\item Average direction function
$Avd_{i}^{t}(\bm{V}(t), \bm{v}_{i}(t)):\mathbb{R}^{d} \times \mathbb{R}^{d} \xrightarrow{} \{ 0, 1 \}$

$\bm{V}(t)$ is the average of $\{ \bm{v}_{1}(t), \bm{v}_{2}(t), ..., \bm{v}_{n}(t)\}$. If an individual's direction of motion deviates from the average by more than a threshold amount $\Theta$ then the individual is in the OFF state: that is, $Avd_{i}^{t}(\bm{V}(t), \bm{v}_{i}(t)) = 1$ when arg($\bm{V}(t)$, $\bm{v}_{i}(t)) \leq \Theta$, and is 0 otherwise. 

\item Centre of mass function
$Com_{i}^{t}(\bm{X}(t), \bm{x}_{i}(t)):\mathbb{R}^{d} \times \mathbb{R}^{d} \xrightarrow{} \{ 0, 1 \}$

$\bm{X}(t)$ is the average of $\{ \bm{x}_{1}(t), \bm{x}_{2}(t), \cdots, \bm{x}_{n}(t)\}$. If an individual is further from $\bm{X}(t)$ than a specified threshold $\Omega$ then the individual is in the OFF state: that is, $Com_{i}^{t}(\bm{X}(t), \bm{x}_{i}(t)) = 1$ when $d(\bm{X}(t)$, $\bm{x}_{i}(t)) \leq \Omega$ and is 0 otherwise.

To obtain the state of the fish school, we take a conjunction of these results to obtain a state for each individual which depends on the pair $(\Theta,\Omega)$:, $s_i(t; \Theta, \Omega) =  Avd_{i}^{t}(\bm{V}(t), \bm{v}_{i}(t)) \wedge Com_{i}^{t}(\bm{X}(t),\bm{x}_{i}(t)) \in \{0,1\}$. The state of the school at time $t$ is then a vector $\bm{s}(t) = (s_1(t), s_2(t), \ldots, s_n(t)) \in \{0,1\}^n$, where the parameter dependence has been omitted for simplicity.

\end{itemize}

\bibliographystyle{unsrt}
\bibliography{references}

\newpage

\section{Supporting Information}

\subsection{Mutual Information}
The mutual information (Eq. ~\ref{MI}) \cite{Oizumi2016} was also calculated in the same manner as $\Phi$ to characterise the susceptibility of $\Phi$. The past and present states of the system are given by the binary variables $X = \{ x_1, x_2, \cdots, x_N \}$ and $Y = \{ y_1, y_2, \cdots, y_N \}$, respectively, where $N$ is the number of elements in the system. The mutual information between the two variables $X$ and $Y$ is expressed by the Kullback-Leibler divergence of the product of their marginal densities from their joint distribution. Intuitively speaking, in our study, the mutual information was used to measure the amount of information shared by the given binary states of fish school over temporal change: $X_t$, $X_{t-\Delta t}$, where $\Delta t = \SI{0.05}{\second}$.

\begin{equation}\label{MI}
\begin{split}
  \min_{q(X_t, X_{t-\Delta t})} D_{KL} [p \left| \right| q] &=  \sum_{X_t,X_{t-\Delta t}} p(X_t,X_{t-\Delta t}) \log{\frac{p(X_t,X_{t-\Delta t})}{p(X_t)p(X_{t-\Delta t})}}  \\
  &= H(X_t) + H(X_{t-\Delta t}) - H(X_{t-\Delta t} \mid X_t)  \\
  &= I(X_t;X_{t-\Delta t})
\end{split}
\end{equation}

\begin{figure}[H]
\includegraphics[width=1.05\linewidth]{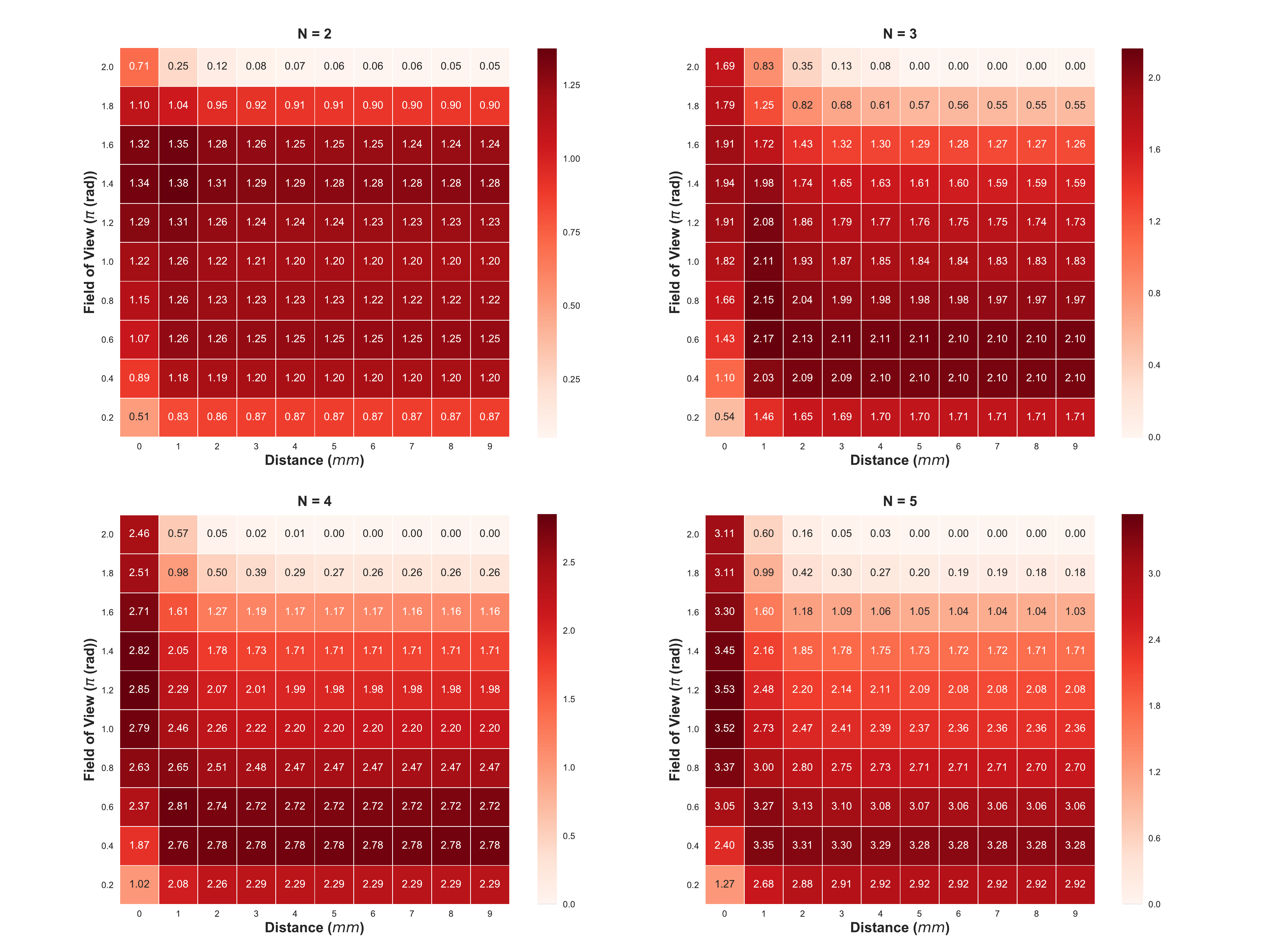}
\caption{{\bf Mutual Information} $I(X;Y)$, as given by the colour bar.}
\label{sfig1}
\end{figure}

The heatmap representations of mutual information values versus various firing thresholds (Fig.~\ref{sfig1}) show different patterns from the heatmap of $\Phi$ values (Fig. 2). First, the results were consistent with the general inequality $0 \leq \Phi \leq I(X;Y)$ shown by Oizumi, \emph{et~al.}\ \cite{Oizumi2016}.  Any peak values of mutual information were higher than the peak values of $\Phi$. Second, the discontinuity between the school size of 3 ($N=3$) and the school size of 4 ($N=4$) was not observed. Third, the exact opposite trends were observed. The values were appeared to be very homogeneous over different settings of firing thresholds. The low and high areas were not so distinguishable as $\Phi$. 

The integrated information $\Phi$ can capture some emergent information dynamics of the fish school which mutual information cannot measure. They essentially have different information: Mutual information or Shannon information is observational and extrinsic, whereas $\Phi$ is causal and intrinsic \cite{Tononi2010}. We saw here this essential differences of the two information quantities.

\subsection{Boids Model}

For comparison, simulated trajectories based on the Boids model (\cite{Couzin2002}) were analysed in the same manner as the trajectories of the real fish school. Boids was developed by Reynolds and the complex and realistic looking behaviour of a group of agents as a whole is determined entirely by local interaction of individual agent choices based on a set of simple rules: Repulsion (Eq. ~\ref{repulsion}), Alignment (Eq. ~\ref{alignment}), and Attraction (Eq. ~\ref{attraction}) . In this study, $N$ agents with position vectors $\bm{x}_i$ and unit direction vectors $\bm{v}_i$ were simulated in continuous two-dimensional space $(3000 \times 2500)$ (the same size as the experimental fish tank). Time was discretised into $t$ computational time steps with a regular spacing $\Delta t = \SI{0.05}{\second}$. When there are $n_r$ agents in the neighbourhood of the agent $i$, the following rules were applied to update the variables of agents at each $t$:

\begin{equation}\label{repulsion}
\bm{d}_r(t + \Delta t) = - \sum_{j \neq i}^{n_r} \frac{\bm{r}_{ij}(t)}{|\bm{r}_{ij}(t)|}
\end{equation}

\begin{equation}\label{alignment}
\bm{d}_o(t + \Delta t) = \sum_{j = 1}^{n_j} \frac{\bm{v}_j(t)}{|\bm{v}_{j}(t)|}
\end{equation}

\begin{equation}\label{attraction}
\bm{d}_a(t + \Delta t) = \sum_{i \neq j}^{n_a} \frac{\bm{r}_{ij}(t)}{|\bm{r}_{ij}(t)|}
\end{equation}

where $n_r = \{ j \mid \bm{r}_{ij} (t) \leq R \}$, $n_o = \{ j \mid \bm{r}_{ij} (t) \leq O \}$, $n_a = \{ j \mid O \leq \bm{r}_{ij} (t) \leq A \}$ and $\bm{r}_{ij} = \frac{(\bm{x}_j-\bm{x}_i)}{|(\bm{x}_j-\bm{x}_i)|}$ is the unit vector in the direction of neighbour $j$. The above rules were summed and averaged with the additive Gaussian noise to determine the trajectories of agents. The update of variables were done synchronously.

\subsubsection{Model parameters}

Parameters are the key to determine the dynamics of the model. In the present study, the model parameters were set to simulate the real experimental data. The average distances were approximately 80 to 140 mm so we set $O = 120~(mm)$ and $R = 10~(mm)$ (= the body length), and $A = \infty$. Thus, the fish school should not part less than 140 mm. This setting was necessary for the swarm to become separated by the boundary conditions. The boundary conditions mimic and reflect the real data. The amplitudes of noise are set to be proportional to the averaged angle change so each agent should have a different noise size.

\begin{table}
\centering
\resizebox{.9\textwidth}{!}{
    \renewcommand{\arraystretch}{1.5}
\begin{tabular}{ccccc}
\hline
    {$N$} & {Average distance \textrm{(mm)}} & {Average velocity \textrm{(mm/s)} } & {Error (S.D.)} & {Minimum distance \textrm{(mm)}}  \\
    \hline
    \multirow{3}{*}{2}  & 166.3  & 11.2 & 0.18 & 1.90  \\
      & 90.67  & 11.32 & 0.23 & 0.10  \\
      & 122.0  & 10.67 & 0.18 & 1.60  \\
    \hline
    \multirow{4}{*}{3}  & 170.8  & 12.55 & 0.23 & 1.80  \\
      & 159.1  & 14.3 & 0.14 & 1.83  \\
      & 173.1  & 12.5 & 0.13 & 2.82  \\
      & 132.0  & 10.0 & 0.19 & 1.67  \\
    \hline
    \multirow{3}{*}{4}  & 164.3  & 11.28 & 0.14 & 1.18  \\
      & 141.5  & 7.95 & 0.12 & 1.38  \\
      & 114.9  & 6.19 & 0.38 & 1.83  \\
    \hline
    \multirow{3}{*}{5}  & 143.8   & 10.83  & 0.28  & 0.79  \\
      & 146.0  & 8.88 & 0.12 & 1.16  \\
      & 143.7  & 10.8 & 0.28 & 1.44  \\
\hline
\end{tabular}
}
\caption{\label{tab:realfishparameters} {\bf Real fish data summary.} $N$: Number of individuals (Unit: None), $x$: Average distance (Unit: \textrm{(mm)}), $v$: Average velocity (Unit: \textrm{(mm)} per second), Error (S.D.) (Unit: Degrees \textrm{(rad)}), $d_{\min}$: Minimum distance (Unit: \textrm{(mm)})}
\end{table}

\begin{table}
\centering
    \renewcommand{\arraystretch}{2} %
    \setlength{\tabcolsep}{11pt}
\begin{tabular}{cccccc}
\hline
    {$N$} & {$R$} & {$O$} & {$A$} & {$v$} & {Error (S.D.)}  \\
\hline
    2  & 10 & 120 & $\infty$ & 11.1 & 0.20  \\
    3  & 10 & 120 & $\infty$ & 12.4 & 0.17   \\
    4  & 10 & 120 & $\infty$ & 8.47 & 0.21  \\
    5  & 10 & 120 & $\infty$ & 10.2 & 0.23   \\ 
\hline
\end{tabular}
\caption{\label{tab:modelparameters} {\bf Summary of model parameters.} The averaged values of $x$, $v$, $\sigma$, $d_{\min}$M for each $N$ were used for the model parameters: $N$ (Unit: None), $x$ (Unit: Units), $v$ (Unit: Units per second), Error (S.D.) (Unit: Degrees (rad)), $d_{\min}$ (Unit: Units).}
\end{table}

\subsubsection{Comparison of trajectories of Real fish and Boid model}

\begin{figure}[hp]
    \centering
    \subfigure[Boid model]
    {
        \includegraphics[width=\linewidth]{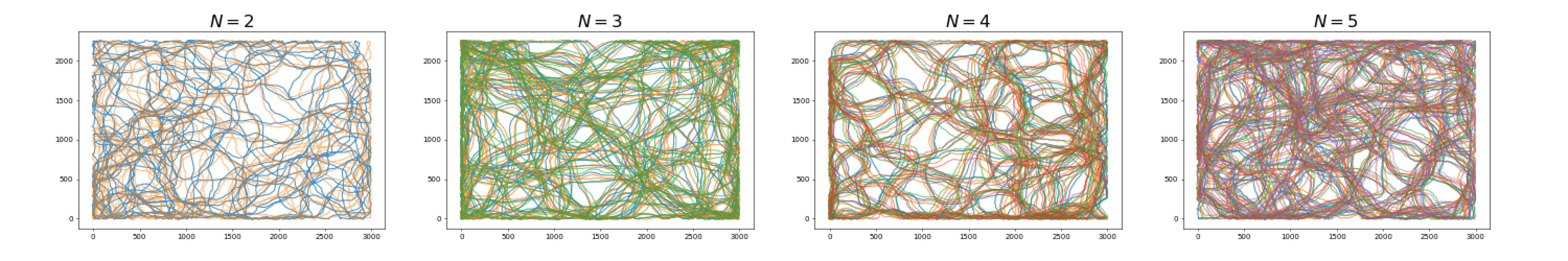}
        \label{fig:sfig21a}
    }
    \\
    \subfigure[Real fish]
    {
        \includegraphics[width=\linewidth]{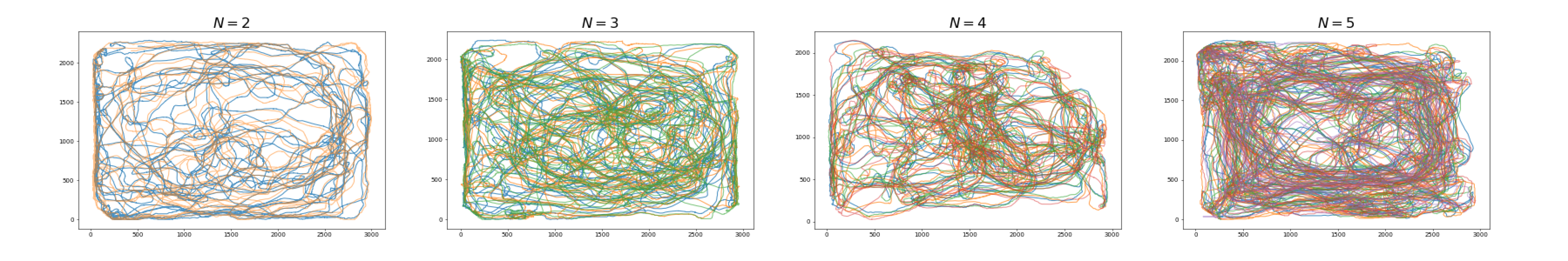}
        \label{fig:sfig21b}
    }
    \caption{{\bf Comparison of Trajectories.} For $T = 20000$ time steps.}
\end{figure}

\begin{figure}[H]
    \centering
    \subfigure[Distance vs. Field of View heatmap]
    {
        \includegraphics[width=.62\linewidth]{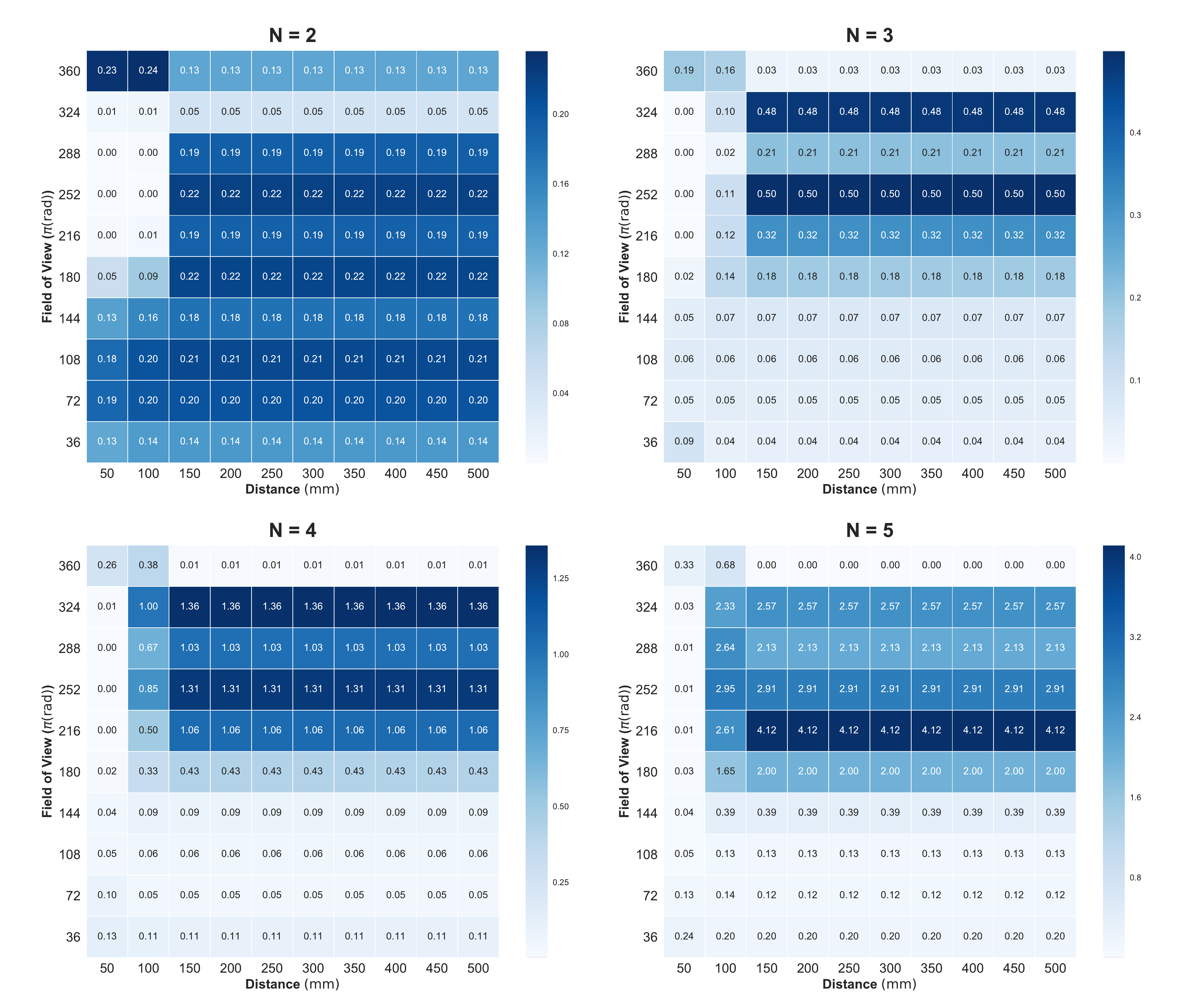}
        \label{fig:sfig22a}
    }
    \\
    \subfigure[The average direction vs. the centre of mass]
    {
        \includegraphics[width=.62\linewidth]{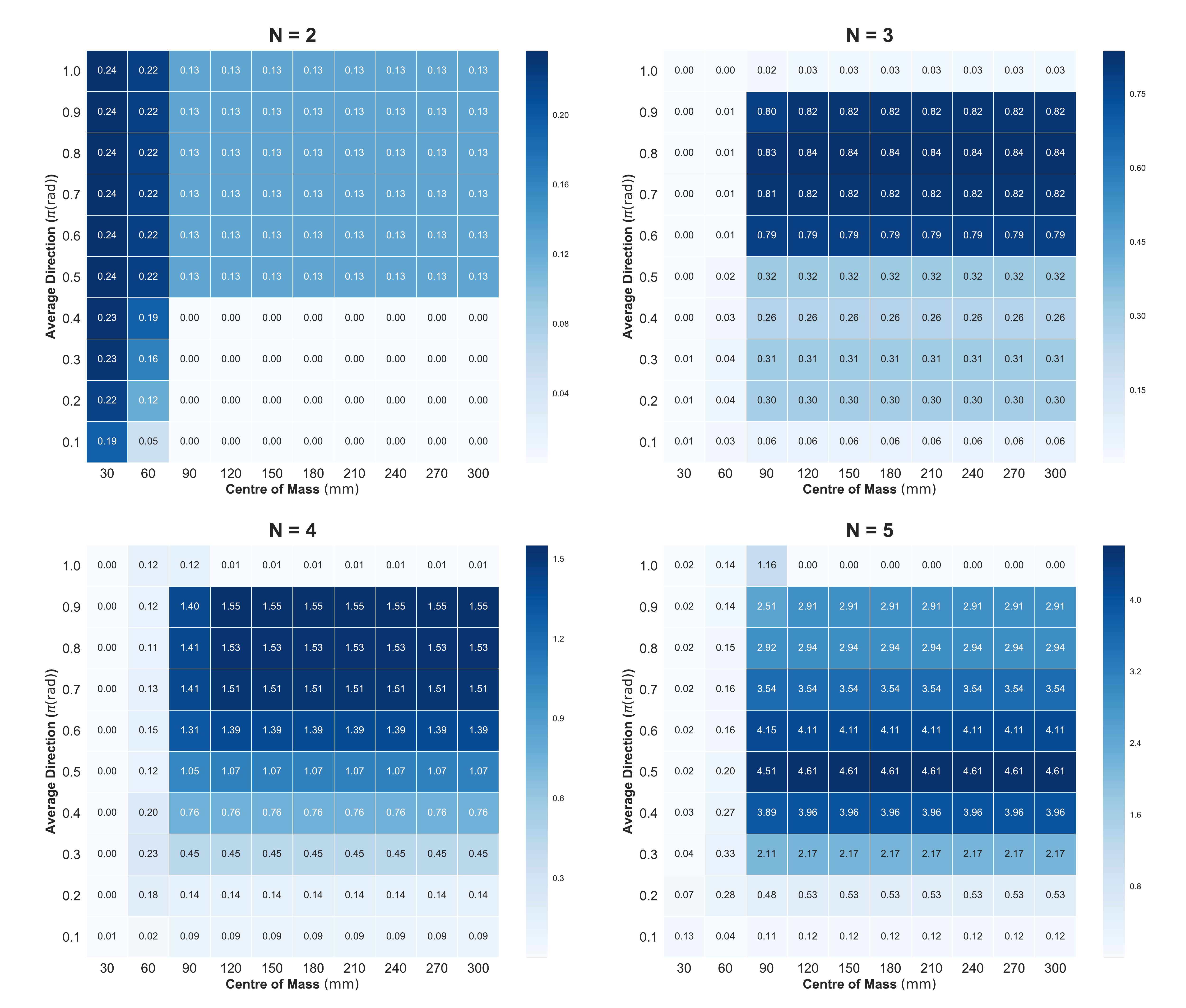}
        \label{fig:sfig22b}
    }
    \caption{{\bf $\Phi$ values of Boids model.}}
    \label{fig:sfig22}
\end{figure}

The constructed boid model (Fig.~\ref{fig:sfig21a}) showed very similar trajectories in terms of the complexity as real fish (Fig.~\ref{fig:sfig21b}). Both Boids and real fish had similar trajectories; however, the Boids' $\Phi$ heatmaps showed different patterns (Fig.~\ref{fig:sfig22a}). The $\Phi$ values of the Boids had large standard deviations. The discontinuity between the $N=3$ and the $N=4$ was not observed. The $N=2$ of Boids were especially different from the $N=2$ of real fish. Similarly, the values of $\Phi$ values for Boids model were large across the different firing setups  (Fig.~\ref{fig:sfig22b}). 

Dynamics of the Boids model and real fish appeared to be very similar; however, when we looked at the $\Phi$ values, those of Boids had large variances, lacked the discontinuity between $N=3$ and $N=4$, and also there were significant differences especially comparing the $N=2$. The Boids $N=2$ had loose and wide distributions of $\Phi$; however, on the other hand, the real $N=2$ had very narrow and susceptible peaks.

\subsection{Analysis of Discontinuity with other parameter settings}

\subsubsection{The enlarged view around Field of View $=2 \pi \ \mathrm{(rad)} $}

\begin{figure}[H]
\includegraphics[width=\linewidth]{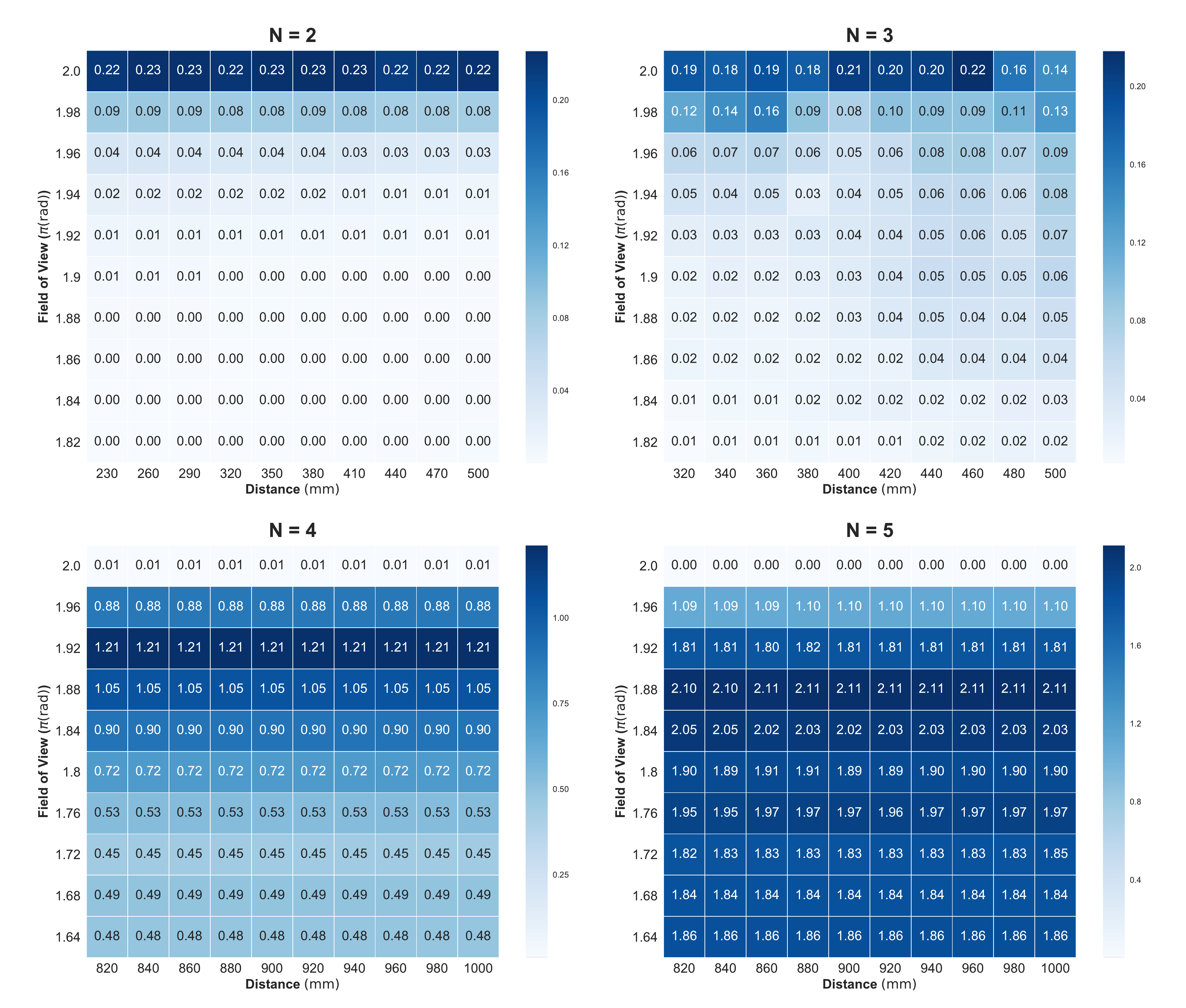}
\caption{{\bf The enlarged view around Field of View $=2 \pi \ \mathrm{(rad)}$.} The $\Phi$ distributions around Field of View = $2 \pi \ rad $ were enlarged to investigate the detailed behaviour of $\Phi$ values. The $\Phi$ for $N=2$ and $N=3$ were discontinued even if we zoomed in at $2 \pi \ rad $. On the other hand, the $\Phi$ values for $N=4$ and $N=5$ around  $2 \pi \ rad $ were almost zero.}
\label{sfig6}
\end{figure}

\newpage

\subsubsection{Other parameter settings}

In addition to Fig. 2, we here visualised $\Phi$ heatmap views for other parameter settings: The Field of View and Turing Rate, and the Distance and Turing Rate.

\begin{figure}[H]
    \centering
    \subfigure[Field of View vs. Turning Rate]
    {
        \includegraphics[width=.63\linewidth]{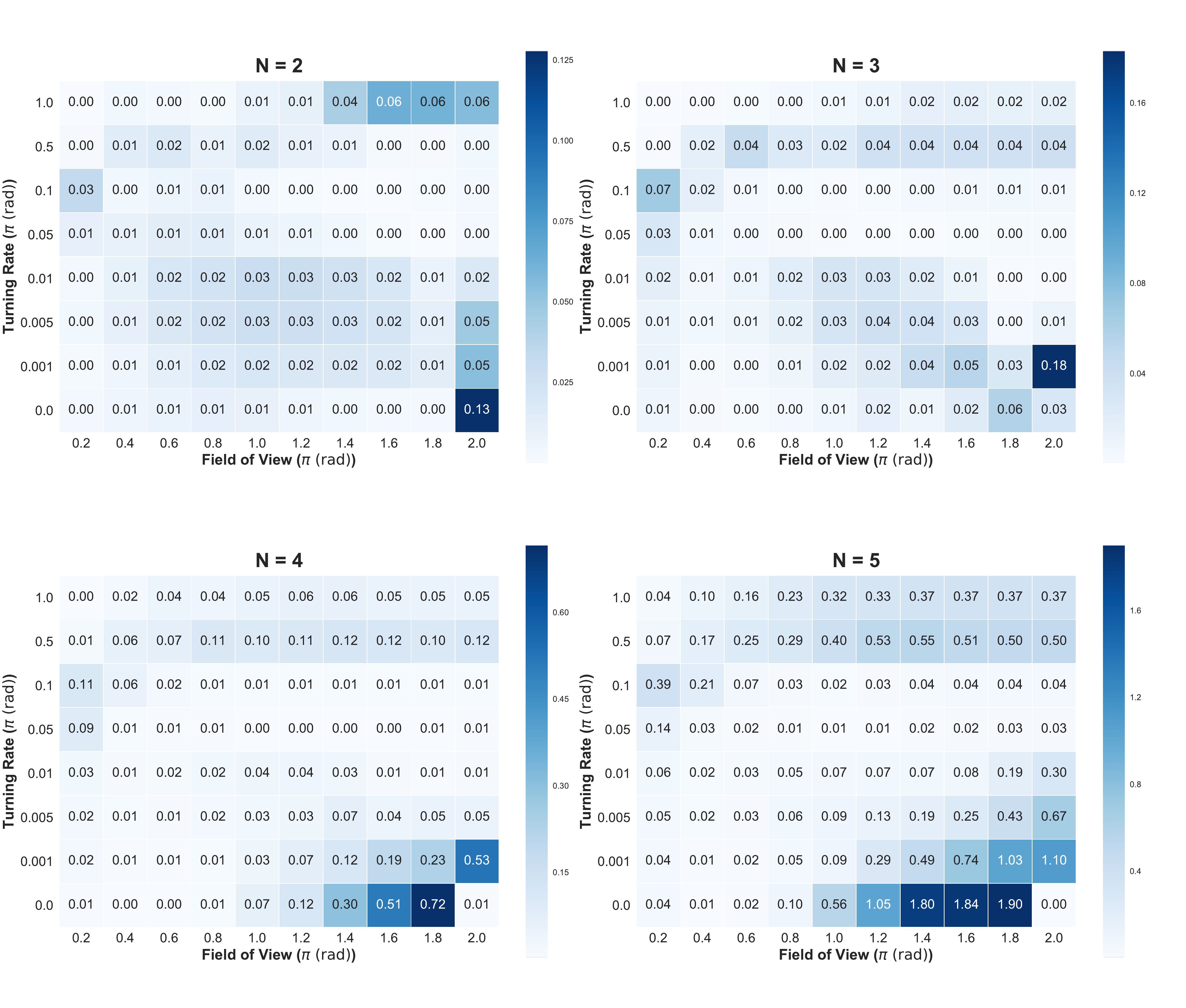}
        \label{fig:sfig32a}
    }
    \\
    \subfigure[Distance vs. Turning Rate]
    {
        \includegraphics[width=.63\linewidth]{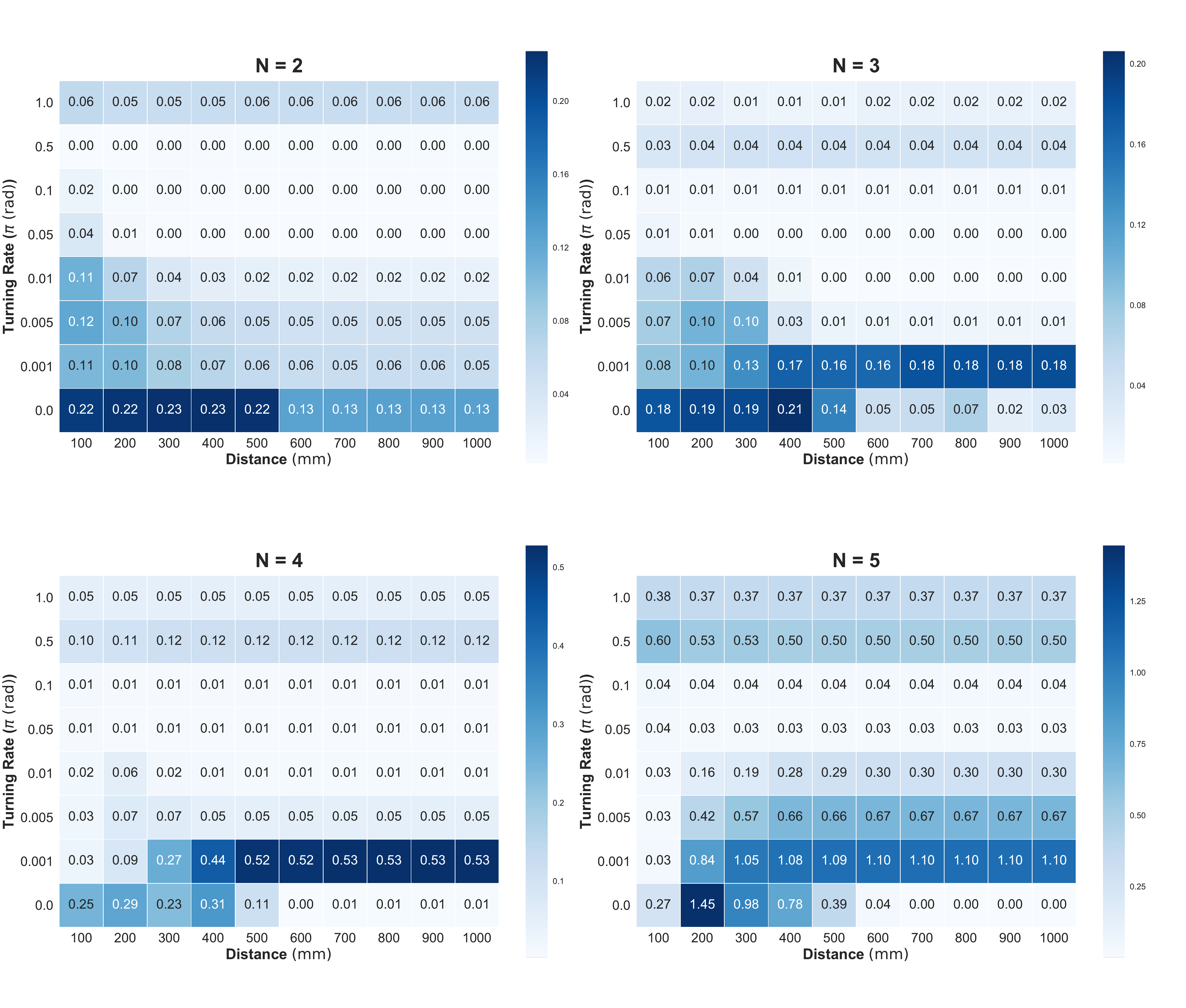}
        \label{fig:sfig32}
    }
    \caption{{\bf Other parameter settings} in the same manner as Fig. 2.}
    \label{fig:sfig32}
\end{figure}

\begin{figure}[H]
    \centering
        \includegraphics[width=.75\linewidth]{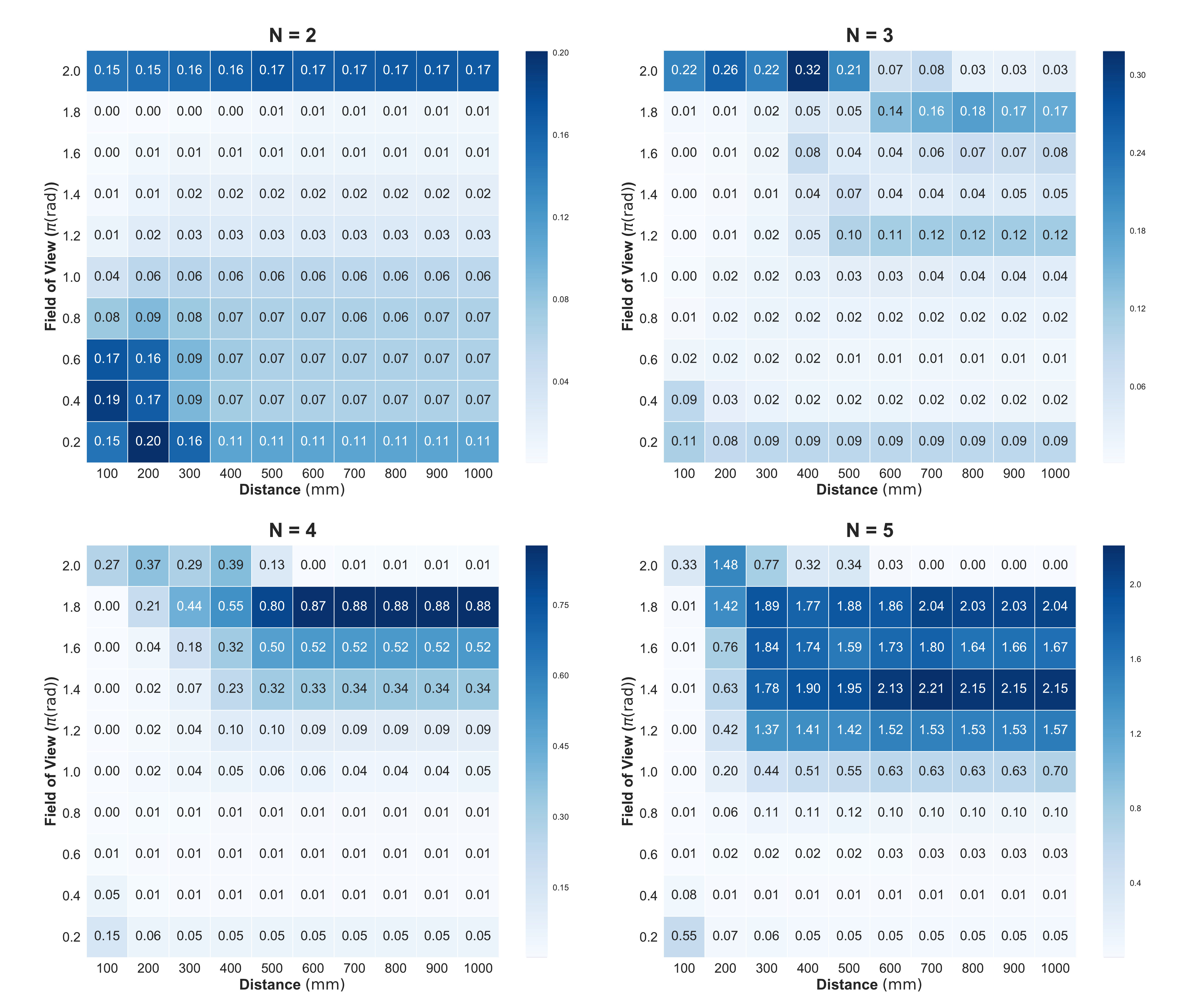}
        \label{fig:sfig33}
    \caption{{\bf Distance vs. Field of View with Time scale $\Delta t = 0.1$.} The most remarkable point here was that when $N=2$ $\Phi$ heatmap of The Distance vs. Field of View showed the followership instead of leadership seen in Fig. 2.}
    \label{fig:sfig33}
\end{figure}

\begin{figure}[H]
    \centering
    \subfigure[Field of View vs. Turning Rate]
    {
        \includegraphics[width=.75\linewidth]{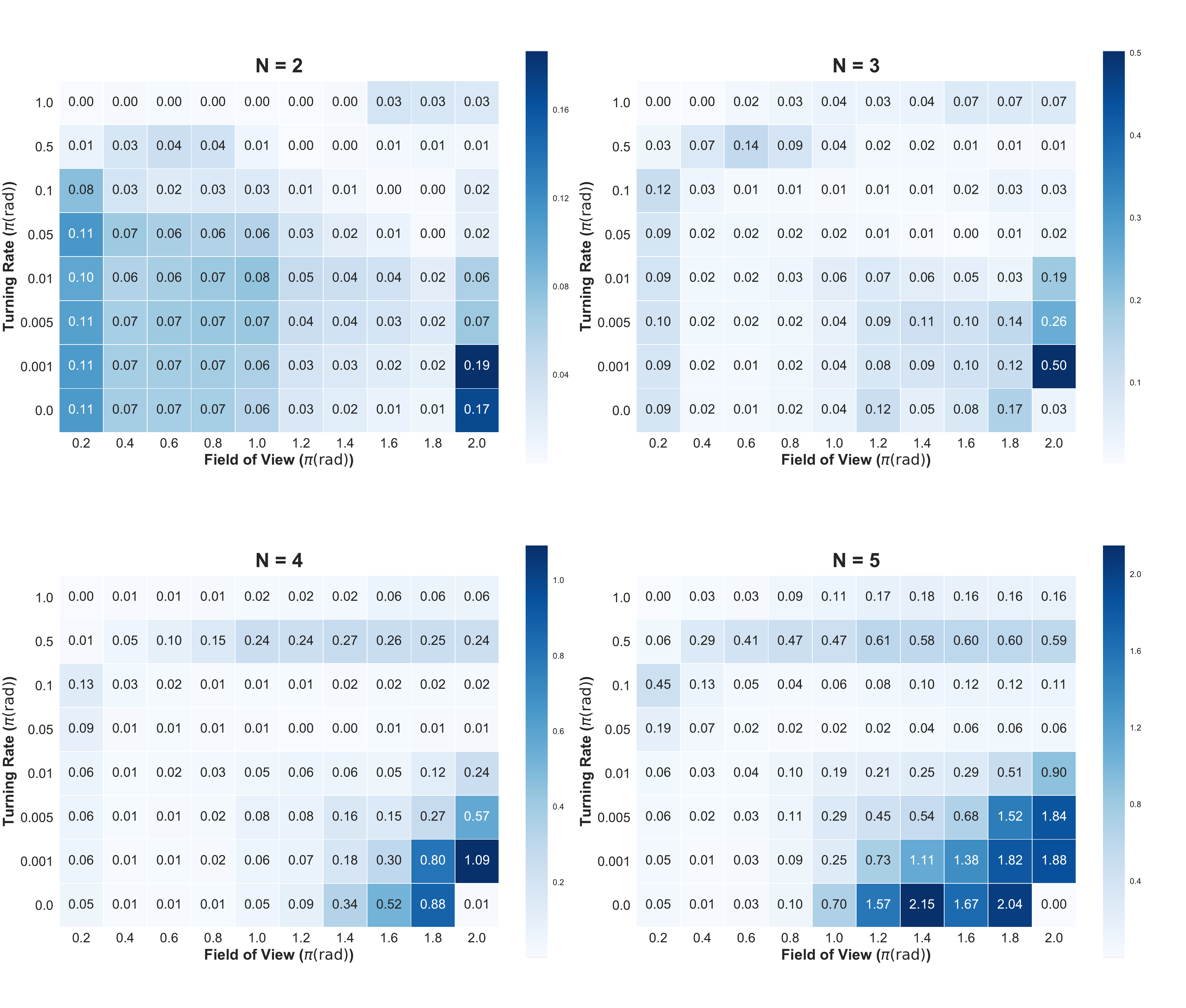}
        \label{fig:sfig34a}
    }
    \\
    \subfigure[Distance vs. Turning Rate]
    {
        \includegraphics[width=.75\linewidth]{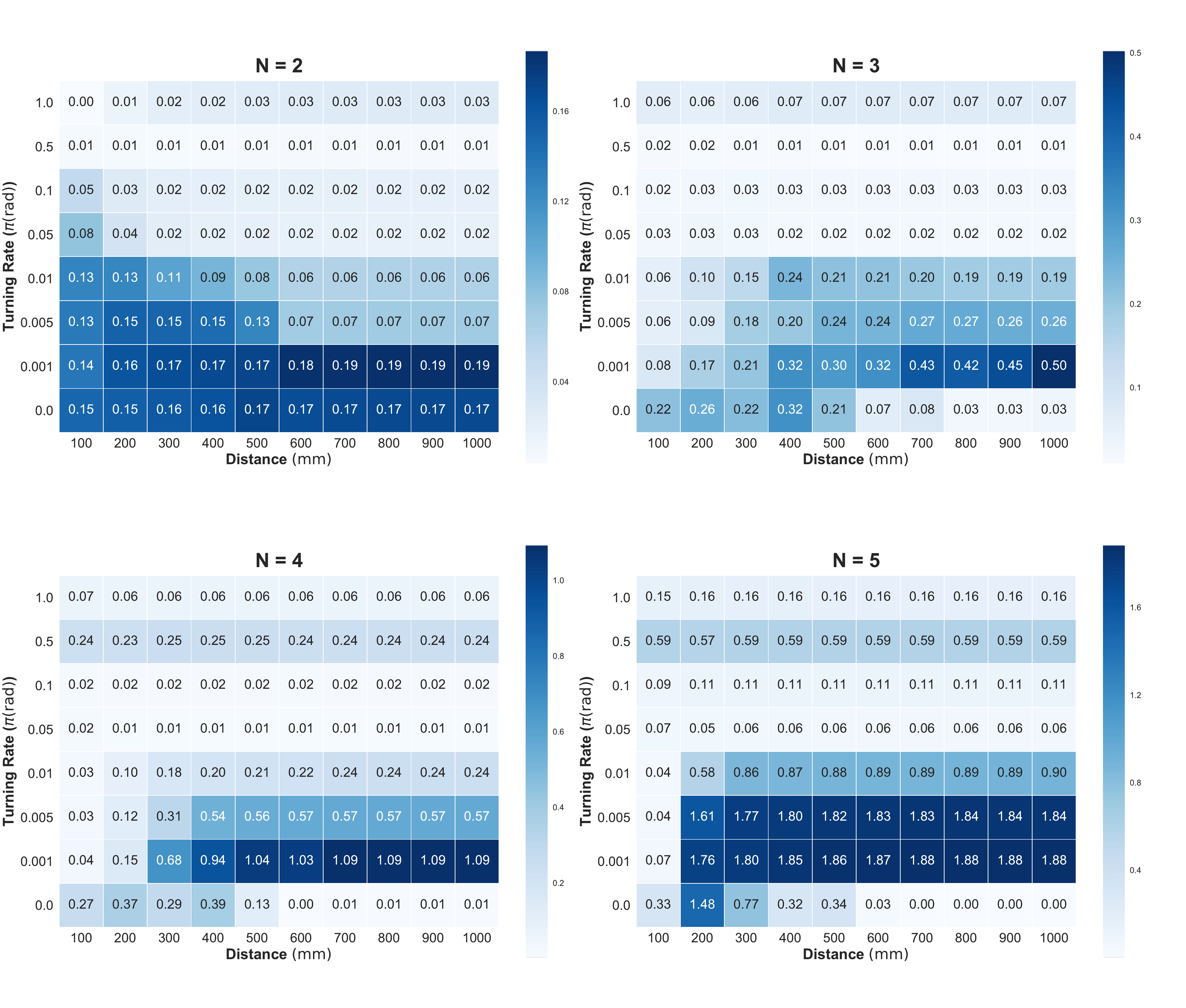}
        \label{fig:sfig34b}
    }
    \caption{{\bf Time scale $\Delta t = 0.1$. with other parameter settings.}}
    \label{fig:sfig34}
\end{figure}

\begin{figure}[H]
    \centering
        \includegraphics[width=.75\linewidth]{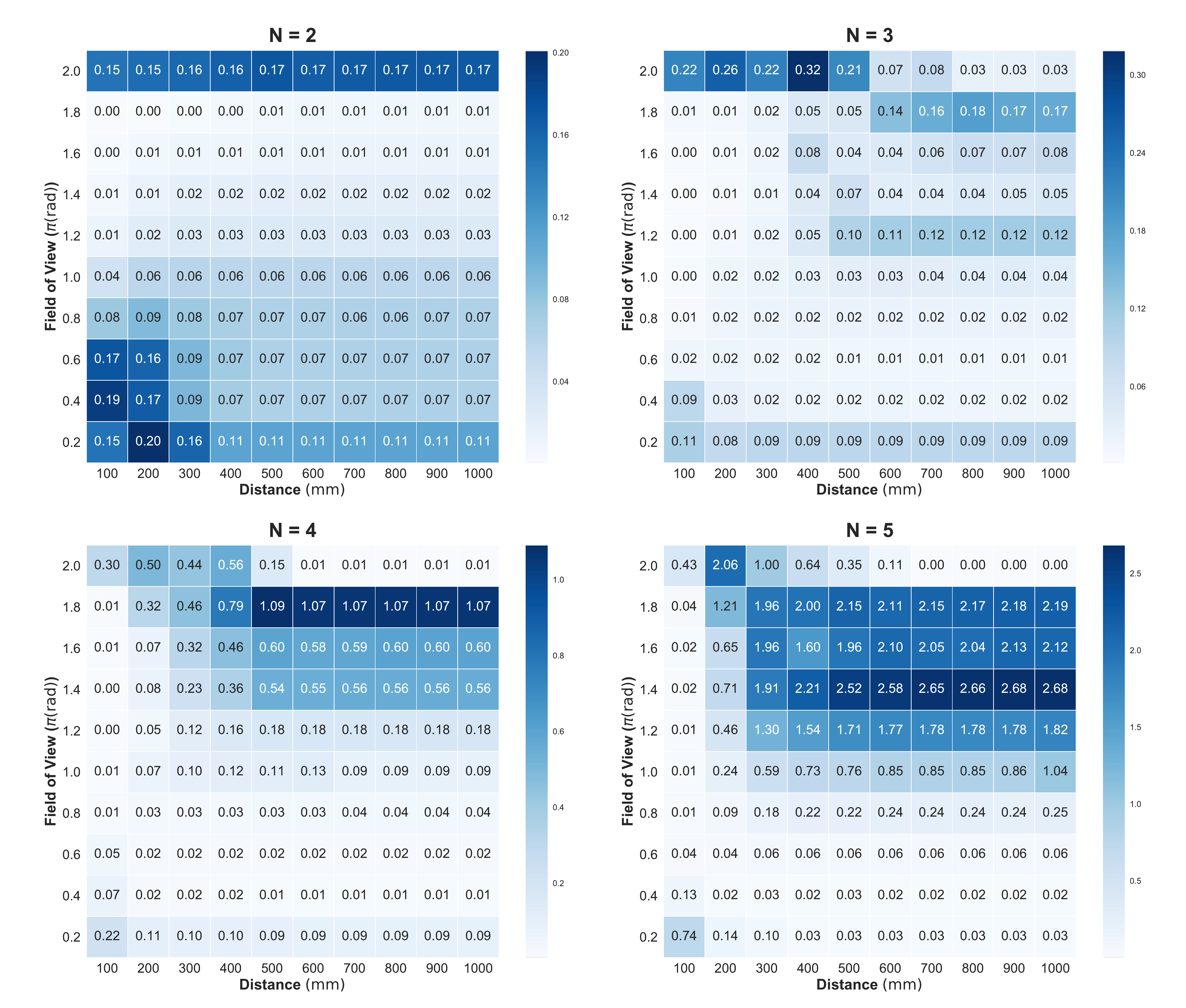}
        \label{fig:sfig33-2}
    \caption{{\bf Distance vs. Field of View with Time scale $\Delta t = 0.2$.} }
    \label{fig:sfig33-2}
\end{figure}

\begin{figure}[H]
    \centering
    \subfigure[Field of View vs. Turning Rate]
    {
        \includegraphics[width=.75\linewidth]{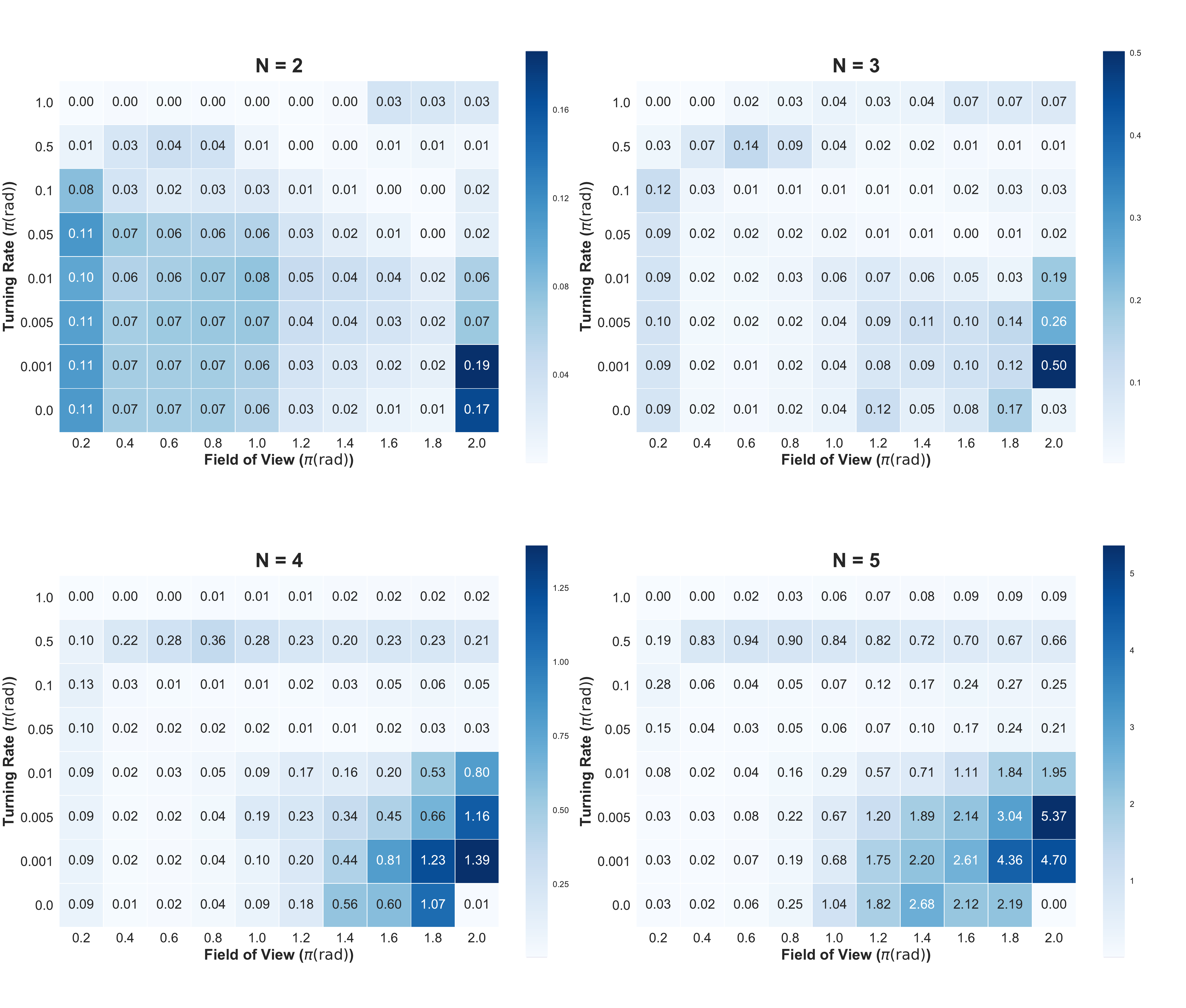}
        \label{fig:sfig34a-2}
    }
    \\
    \subfigure[Distance vs. Turning Rate]
    {
        \includegraphics[width=.75\linewidth]{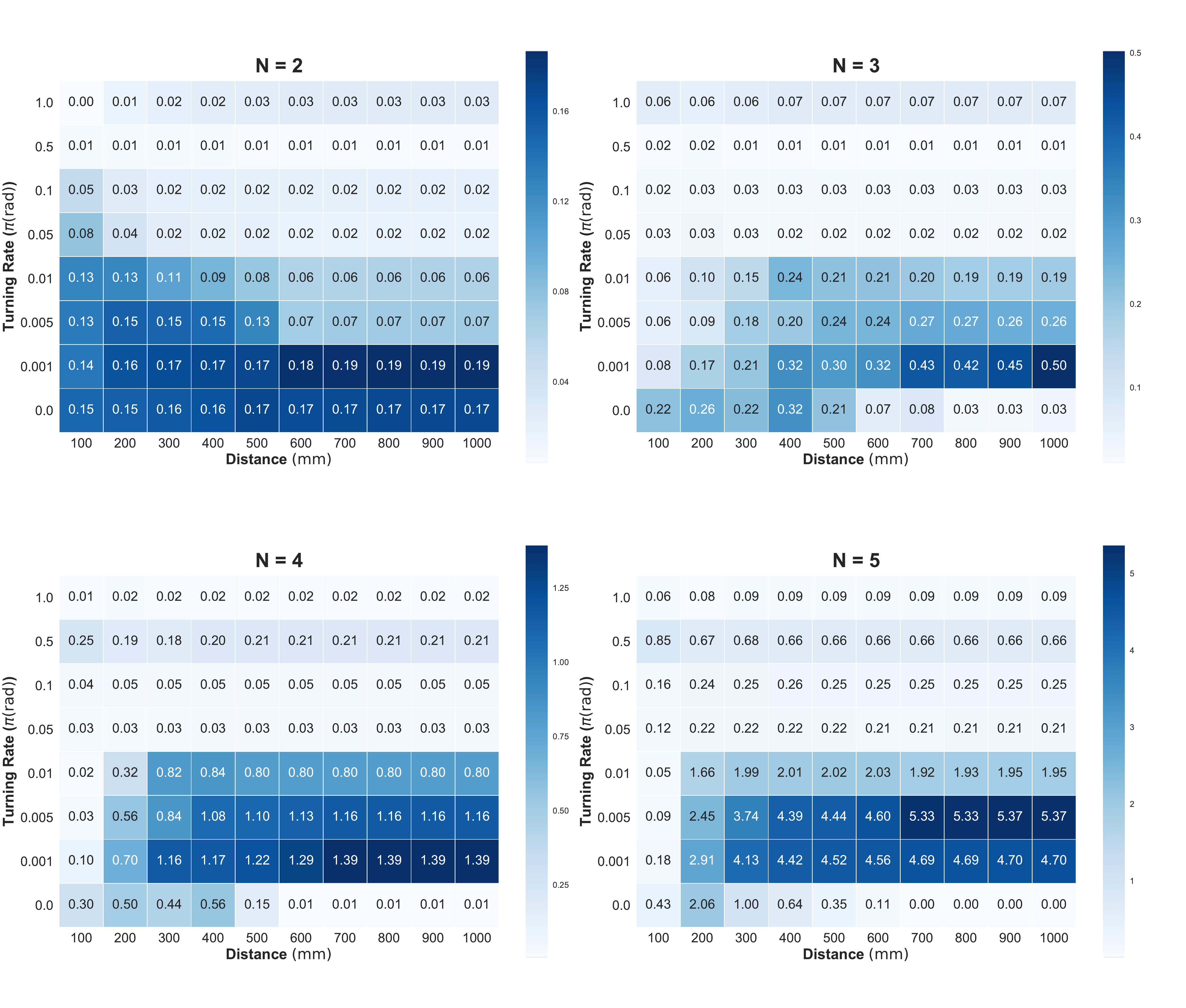}
        \label{fig:sfig34b-2}
    }
    \caption{{\bf Time scale $\Delta t = 0.2$. with other parameter settings.}}
    \label{fig:sfig34-2}
\end{figure}

\subsubsection{Concept}
\begin{figure}[H]
    \centering
    \includegraphics[width=.9\linewidth]{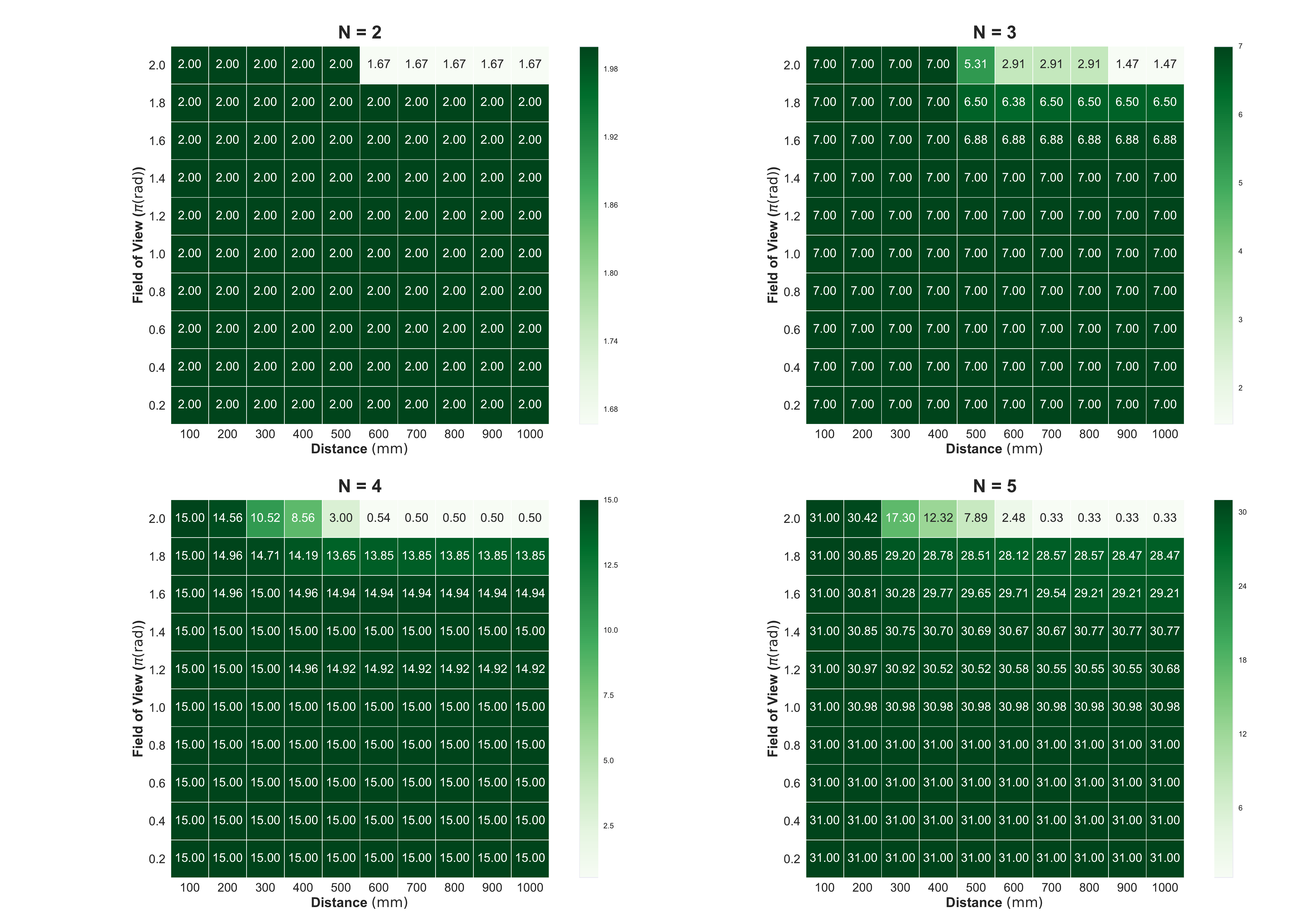}
    \label{fig:sfig35}
    \caption{{\bf The Number of Concepts: Mechanisms that specify maximally irreducible cause and effect (MICE) repertoires:} Distance vs. Field of View. "PyPhi.Subsystem.concept()" was used to comupute Concepts. The number of concepts were assessed here. Some of parameter settings have a large number of concepts; the $\Phi$ values still remain low. This perhaps implies a potential of having larger $\Phi$'s.}
    \label{fig:sfig35}
\end{figure}

\begin{figure}[H]
    \centering
    \subfigure[Field of View vs. Turning Rate]
    {
        \includegraphics[width=.9\linewidth]{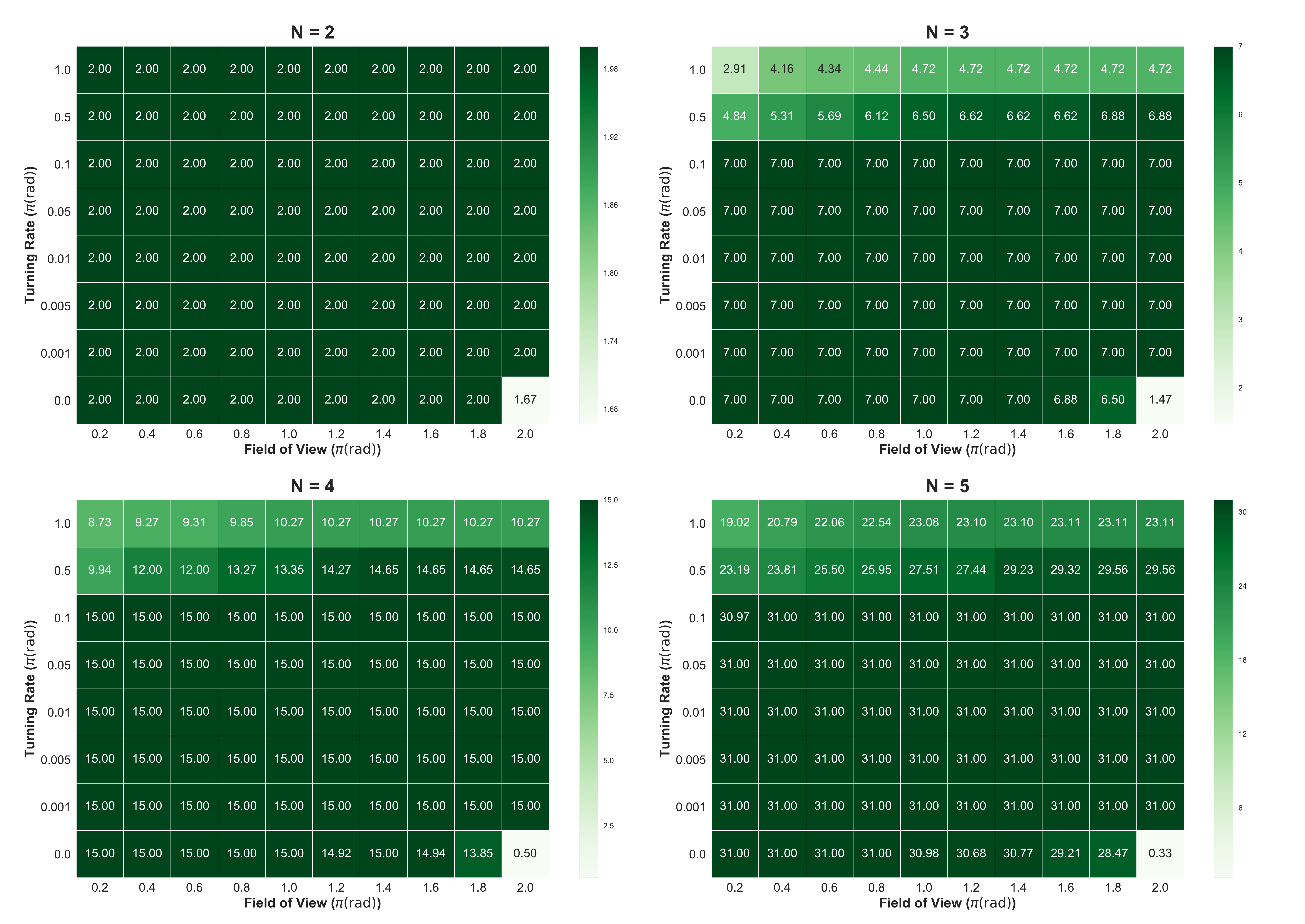}
        \label{fig:sfig36a}
    }
    \\
    \subfigure[Distance vs. Turning Rate]
    {
        \includegraphics[width=.9\linewidth]{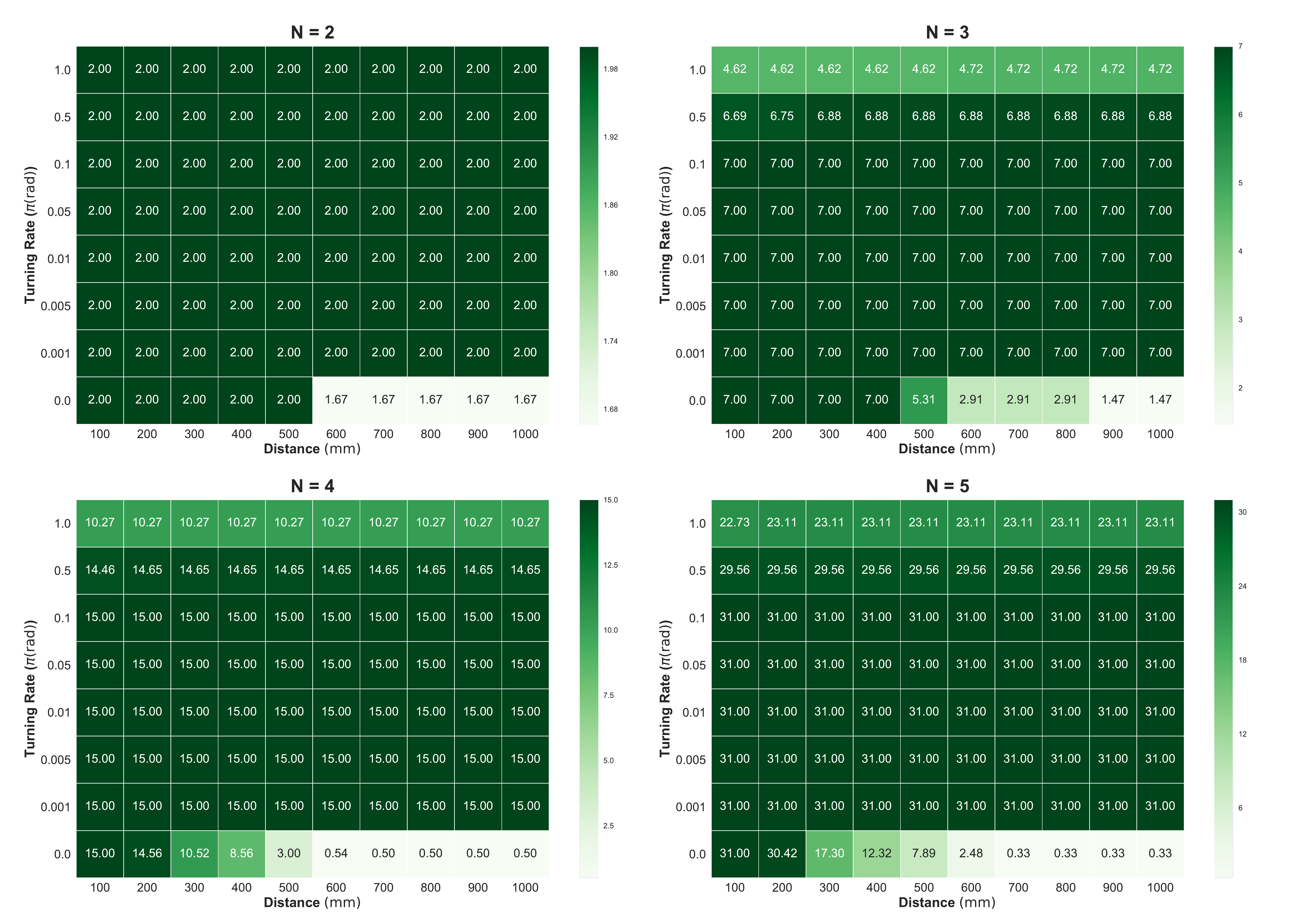}
        \label{fig:sfig36b}
    }
    \caption{{\bf The Number of Concepts: Mechanisms that specify maximally irreducible cause and effect (MICE) repertoires:} Other parameter settings. "PyPhi.Subsystem.concept()" was used to comupute Concepts.}
    \label{fig:sfig36}
\end{figure}

\subsubsection{Susceptibility}

Susceptibility is the measure of fluctuations or the response of an extensive property such as the order parameter to a small external perturbation to give variation of an intensive property. Power-law divergences of quantities like the magnetic susceptibility in the ferromagnetic phase transition in critical phenomena are well-known. The magnetic susceptibility per spin of Ising model (Eq. ~\ref{magnetic_susceptibility}) is defined as the derivative of the average total magnetisation with respect to the external field at fixed temperature and related to the variance of the average total magnetisation through the fluctuation-dissipation theorem,

\begin{equation}\label{magnetic_susceptibility}
\begin{split}
\chi :&= \frac{1}{N}\left(\frac{\partial \langle M \rangle}{\partial H}\right)_T \\
&= \frac{k_BT}{N}\left(\langle M^2 \rangle - \langle M \rangle^2\right)
\end{split}
\end{equation}

In the recent work by Khajehabdollahi \cite{Khajehabdollahi_2018}, $\Phi$ of the small Ising model was interpreted as an order parameter and the author discovered that the critical temperature maximises the susceptibility of $\Phi$ (Eq. ~\ref{phi_susceptibility}).

\begin{equation}\label{phi_susceptibility}
\sigma^2(\Phi) = \langle \Phi^2 \rangle- \langle \Phi \rangle^2
\end{equation}

$\Phi$ susceptibility of real fish school was assessed here. The large susceptibilities of $\Phi$ are mostly corresponding to the large $\Phi$ values; however, it is interesting to see they do not simply correspond to each other.

\begin{figure}[H]
    \centering
    \includegraphics[width=.9\linewidth]{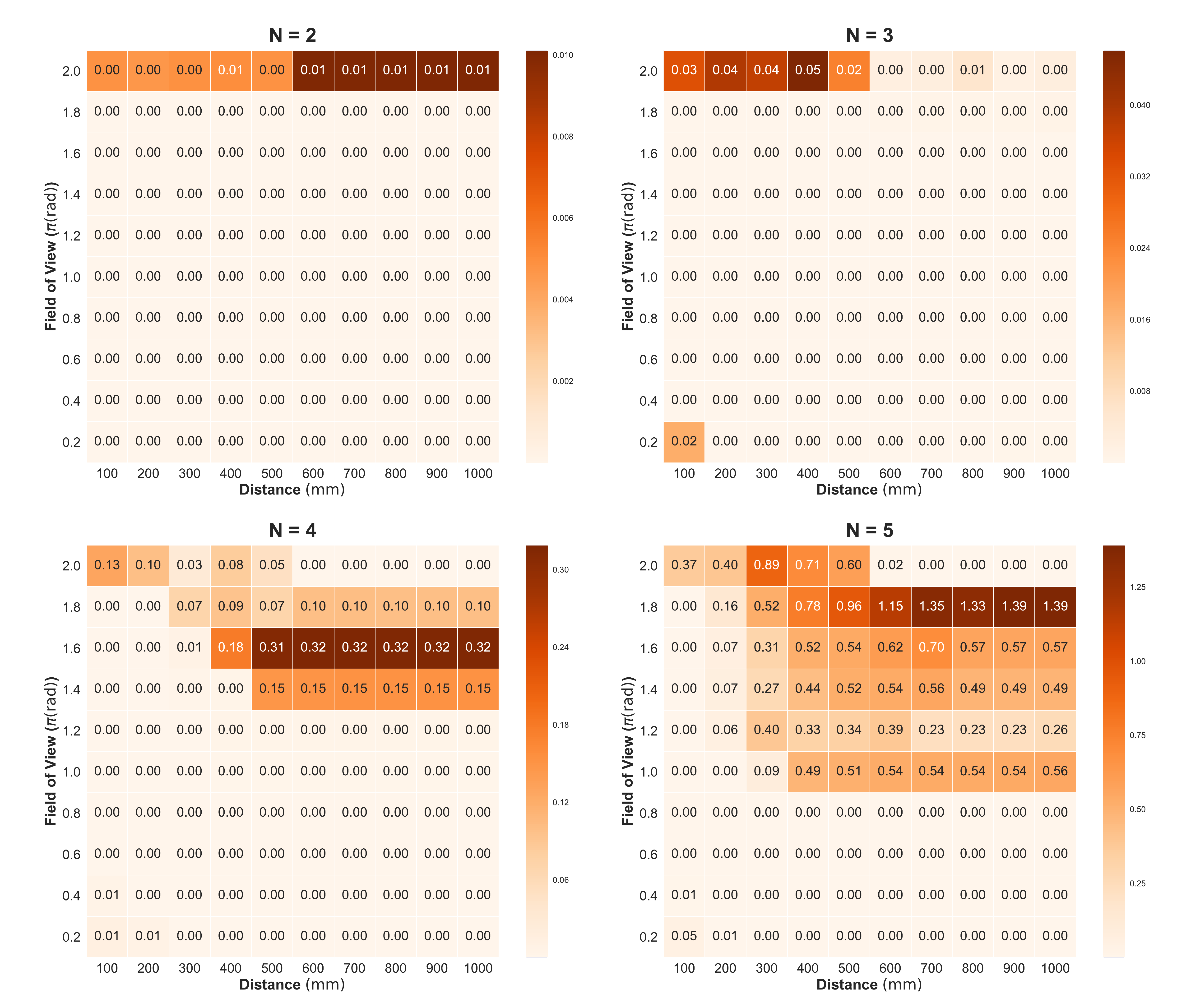}
    \label{fig:sfig37}
    \caption{{\bf Susceptibility: Distance vs. Field of View.}} 
    \label{fig:sfig37}
\end{figure}

\begin{figure}[H]
    \centering
    \subfigure[Field of View vs. Turning Rate]
    {
        \includegraphics[width=.75\linewidth]{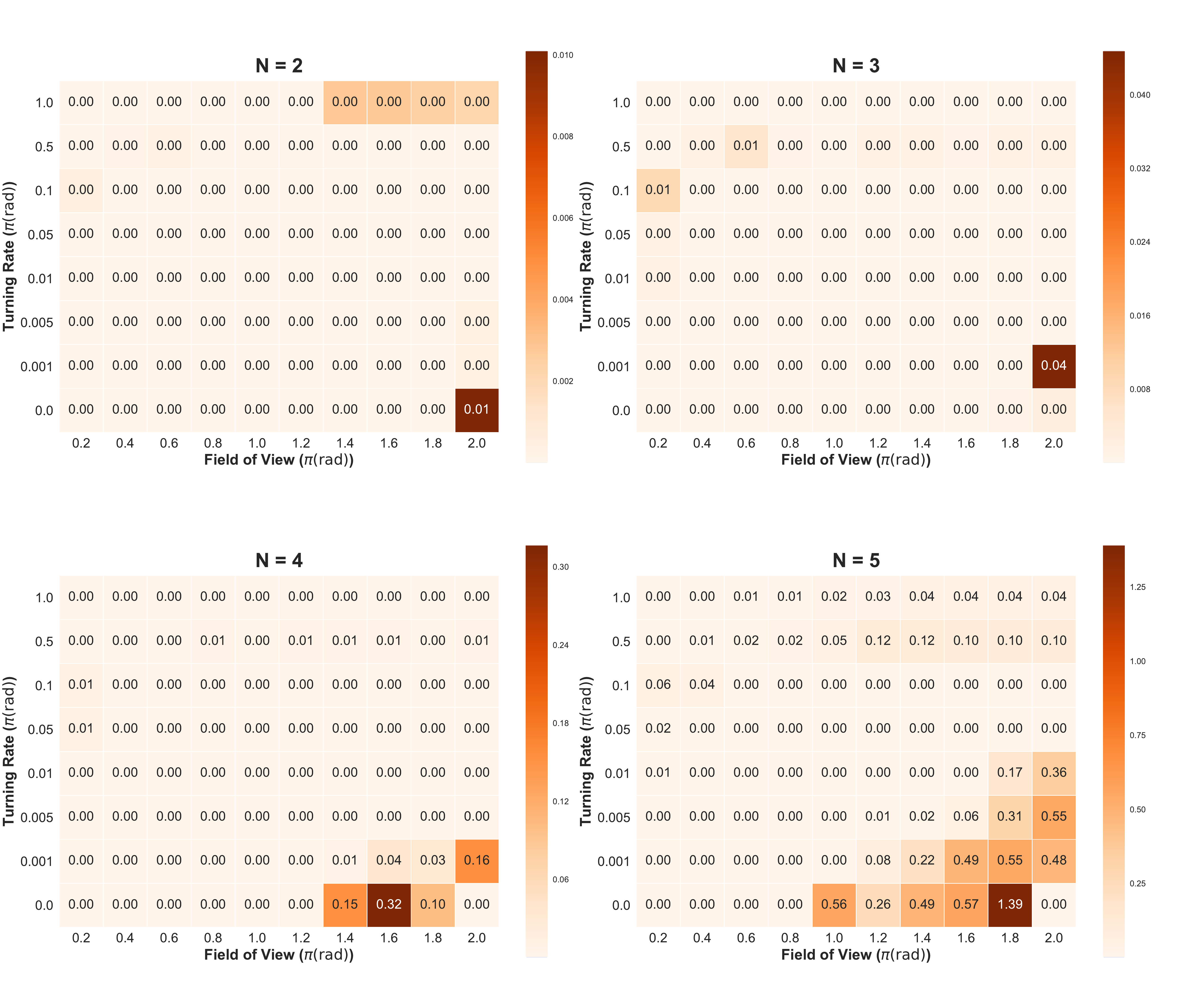}
        \label{fig:sfig38a}
    }
    \subfigure[Distance vs. Turning Rate]
    {
        \includegraphics[width=.75\linewidth]{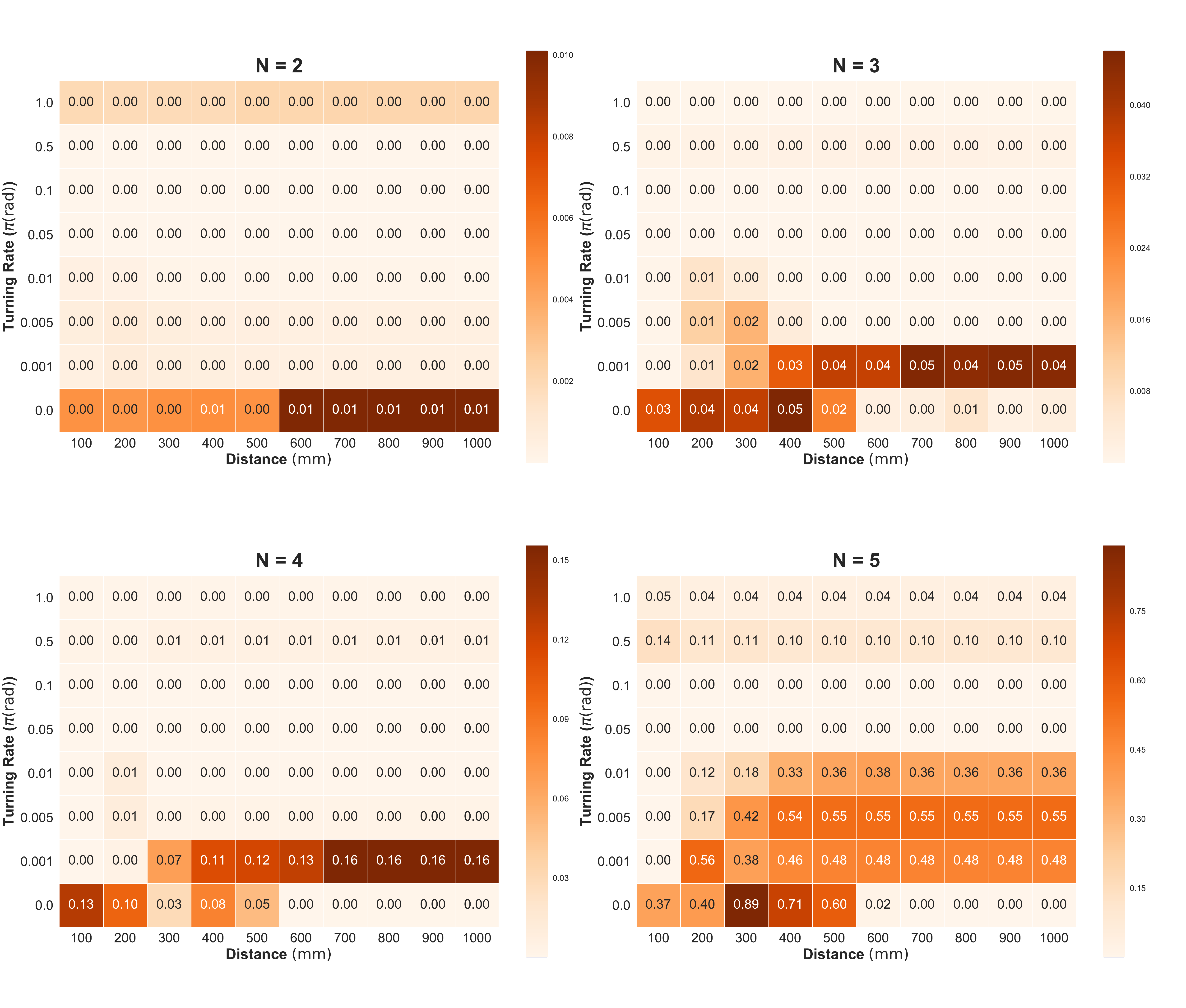}
        \label{fig:sfig38b}
    }
    \caption{{\bf Susceptibility with other parameter settings.} }
    \label{fig:sfig38}
\end{figure}

\subsection{$\Phi(N)$ increase with the group size $N$}

The mean values and the standard deviations of integrate information $\Phi$ increase as the size of fish school $N$ increases.

\begin{figure}[H]
\includegraphics[width=\linewidth]{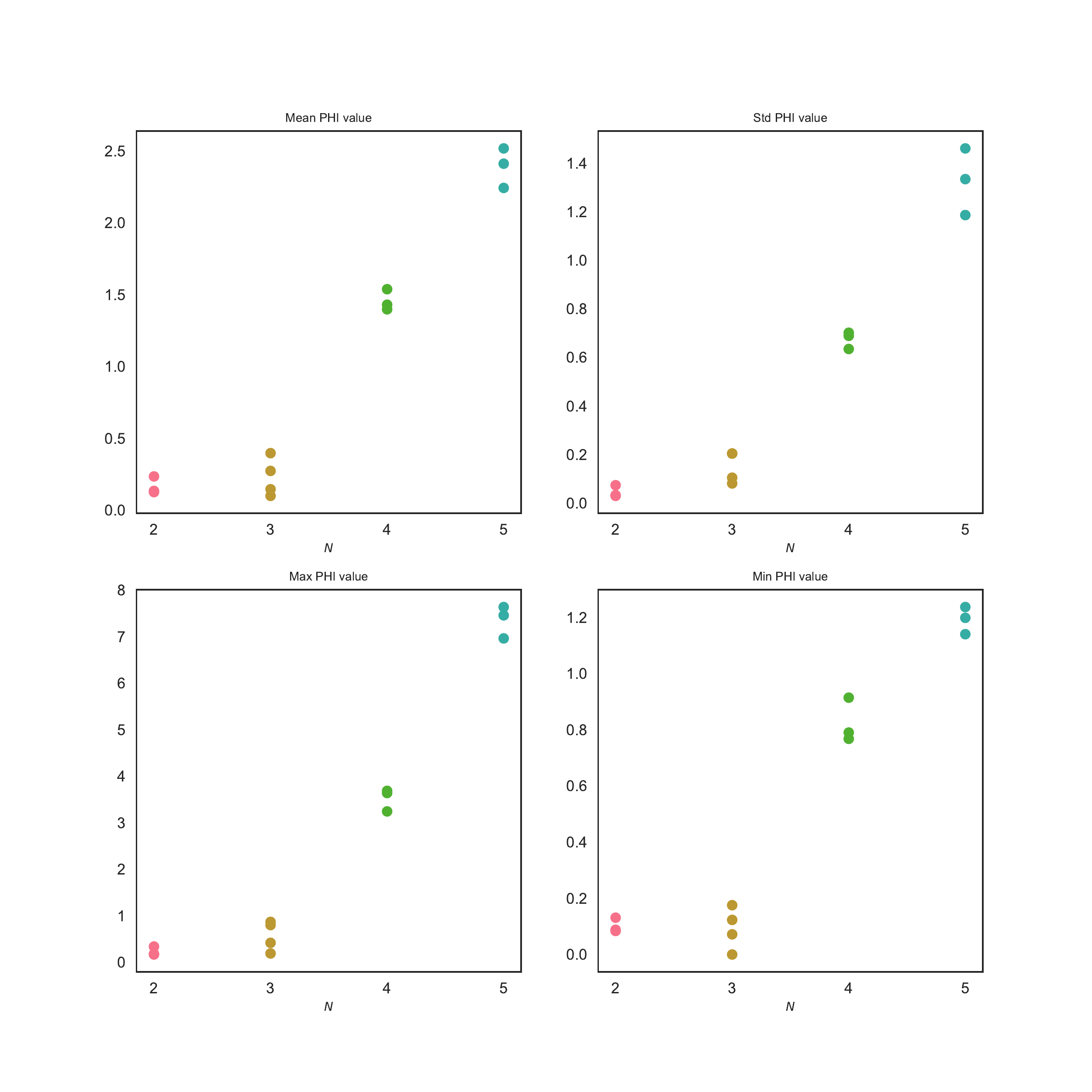}
\caption{Mean Values $\bar{\Phi}_N$ , Standard deviation $\sigma(\Phi)_N$, Max Values $\max\{\Phi\}_N$, and Min Values $\min\{\Phi\}_N$. The value was calculated for each group. Groups $N=2$, $N=4$,$N=5$ have 3 samples and Group $N=3$ has 4 samples. Parameters were picked up from the peak values of $\Phi$-dist map}
\label{sfig39}
\end{figure}

\newpage

\subsection{A short summary of integrated information $\Phi$}

Integrated information theory models a system $S$ by a discrete time multivariate stochastic process:
\begin{equation}\label{model}
p(X_0,X_{\Delta t},\ldots,X_t,X_{t +\Delta t},\ldots, X_T)
\end{equation}
that fulfils the Markov property:
\begin{equation}\label{markov}
p(X_0,X_{\Delta t},\ldots,X_t,X_{t +\Delta t},\ldots, X_T) = p(X_0)\prod_{t=\Delta t}^{T}p(X_t \mid X_{t-\Delta t})
\end{equation}
Such a discrete dynamical system $S$ is defined by a directed graph of interconnected nodes (in this study we assumed a complete graph.) and its transition probability matrix (TPM). The TPM specifies the conditional probability distribution $p(X_t \mid X_{t-\Delta t})$. Each state vector $X_t$ comprises binary variables $x_{t_i}$,$i=1,2,\ldots,n$($n \in N$).

A joint distribution $p_{\mathrm{casue-effect}}$ is defined
\begin{equation}\label{cause-effect}
p_{\mathrm{cause-effect}}(X_{t-\Delta t},X_t) := p_u(X_{t-\Delta t})p_{\mathrm{effect}}(X_t\mid X_{t-\Delta t}) 
\end{equation}

The marginal distribution $p_u(X_{t-\Delta t})$ is an uniform distribution to give the maximum entropy distribution.

From the joint probability above

the backward transitional probability distribution

\begin{equation}\label{effect}
p_{\mathrm{effect}}(X_{t-\Delta t}\mid X_t) := \frac{p_{\mathrm{casue-effect}}(X_{t-\Delta t},X_t)}{\sum_{X_{t-\Delta t}}p_{\mathrm{casue-effect}}(X_{t-\Delta t},X_t)}
\end{equation}

and the forward transitional probability distribution

\begin{equation}\label{cause}
p_{\mathrm{cause}}(X_t\mid X_{t-\Delta t}) := p((X_t\mid X_{t-\Delta t}))
\end{equation}

are constructed and referred to as the {\it cause repertoire} and the {\it effect repertoire} of state $X_t$, respectively. The {\it cause repertoire} and the {\it effect repertoire} are calculated for a set of nodes within the subsystem, or a {\it mechanism} $M \subseteq S$, over another set of nodes within the subsystem, or a {\it purview} of the mechanism.

After assessing the information of a mechanism over a purview we next consider its {\it integrated information} $\phi_{\mathrm{cause-effect}}$ of a set of system elements in a state X defined as

\begin{equation}\label{phi_ce}
\phi_{\mathrm{cause-effect}}:= \min\{ \phi_{\mathrm{effect}},  \phi_{\mathrm{cause}} \}
\end{equation}

\begin{equation}\label{phi_e}
\phi_{\mathrm{effect}}:= \min_{i \in I} \{ D \left(  p_{\mathrm{effect}} \left| \right| p_{\mathrm{effect}}^{(i)}  \right)  \}
\end{equation}

\begin{equation}\label{phi_c}
\phi_{\mathrm{cause}}:= \min_{i \in I} \{ D \left( p_{\mathrm{cause}} \left| \right| p_{\mathrm{cause}}^{(i)}  \right)  \}
\end{equation}

where the system is decomposed by all possible ways into $I$.

The integrated information $\phi$ is assessed by quantifying the extent to which the cause and effect repertoires of the mechanism-purview pair can be reduced to the repertoires of its parts. The amount of irreducibility of a mechanism over a purview with respect to a partition is quantified as the divergence between the unpartioned repertoire $p$ and the partitioned repertoire $p^{(i)}$. The partition that yields the minimum irreducibility is called the {\it minimum-information partition} (MIP). The integrated information $\phi$ of a mechanism-purview pair is defined as the divergence between the unpartitioned repertoire and the repertoires partitioned by MIP.
The maximum $\phi$ value is then searched over all possible purviews to find {\it maximally-irreducible cause} (MIC) and {\it maximally-irreducible effect} (MIE) specified by a mechanism.

\begin{equation}\label{maxphi}
\phi_{\mathrm{cause}}^{\max}:=\max_{j\in C}\{\phi_{\mathrm{cause}}^j\},\phi_{\mathrm{effect}}^{\max}:=\max_{j\in C}\{\phi_{\mathrm{effect}}^j\}
\end{equation}

where $C=2^N-1$ (In this paper we adopted a "cut one" approximation which only evaluates $2N$ bipartitions severing the edges from a single node to the rest of the network.).

The $\phi$ value of the concept as a whole or the maximally integrated cause-effect information is the minimum of maximally integrated cause information $\phi_{\mathrm{cause}}$ and maximally integrated effect information $\phi_{\mathrm{effect}}$.

\begin{equation}\label{minphi}
\phi_{\mathrm{cause-effect}}^{\max} := \min \{ \phi_{\mathrm{cause}}^{\max}, \ \phi_{\mathrm{effect}}^{\max}   \}
\end{equation}

If the mechanism's MIC has $\phi_{\mathrm{cause}}>0$ and its MIE has $\phi_{\mathrm{effect}}>0$, (equivalently $\phi_{\mathrm{cause-effect}}^{\max}>0$)  then the mechanism is said to specify a {\it concept}.

We then compute the {\it cause-effect structure} (CES), the set of all concepts specified by the subsystem characterising all of the causal constraints intrinsic to the physical system, by simply iterating the computation of concepts over all mechanisms $M \in \mathcal{P(S)}$, where $\mathcal{P(S)}$ is the power set of subsystem nodes. 

Integrated conceptual information $\Phi$ ({\it Big Phi}), a measure of the system's strong/integration irreducibility, is assessed by partitioning the set of elements into subsets with unidirectional cuts. Unidirectional bipartitions $P_{\rightarrow} = \{ S^{(1)}; S^{(2)} \}$ of the physical system $S$ are performed by partitioning the subsystem into two parts $S^{(1)}$ and $S^{(2)}$ and cutting the edges going from one part $S^{(1)}$ to another $S^{(2)}$ (the connections are substituted by noise). We then calculate the CES of the partitioned system $C(S^{P_{\rightarrow}})$ and compare it to $C(S)$ to evaluate the difference made by the partition. MIP, a search over all possible directed partitions is then performed to identify the one that makes the least difference to the CES. Integrated information $\Phi$ ({\it Big Phi}) measures the irreducibility of a cause-effect structure, by quantifying the difference the MIP makes to the concepts and their $\phi$ values of the system.

\begin{equation}\label{bigphi}
\Phi = \min_{P_{\rightarrow}} D \left( C(S), C(S^{P_{\rightarrow}}) \right)
\end{equation}

The difference $D$ between two cause-effect structures is evaluated by an extended version of the Earth Mover's Distance: the cost of transforming one cause-effect structure $C(S)$ into another $C(S^{P_{\rightarrow}})$ in concept space.

\end{document}